\documentclass[lettersize,journal]{IEEEtran}
\IEEEoverridecommandlockouts
\usepackage{cite}
\usepackage{citesort}
\usepackage{amsmath,amsfonts,amsthm,amssymb}
\allowdisplaybreaks
\usepackage{bbm}
\usepackage{algpseudocode}
\usepackage{algorithm}
\usepackage{graphicx}
\usepackage{subfigure}
\graphicspath{{figure/}}
\usepackage{textcomp}
\usepackage{booktabs}
\usepackage{setspace}
\usepackage{tabularx}
\newcommand{\ra}[1]{\renewcommand{\arraystretch}{#1}}

\newcolumntype{L}{>{\hspace*{-\tabcolsep}}l}
\newcolumntype{R}{c<{\hspace*{-\tabcolsep}}}
\usepackage[table]{xcolor}
\usepackage{xcolor}
\usepackage{extarrows}
\usepackage{times}
\usepackage[justification = centering]{caption}
\definecolor{lightblue}{rgb}{0.93,0.95,1.0}

\def\BibTeX{{\rm B\kern-.05em{\sc i\kern-.025em b}\kern-.08em
		T\kern-.1667em\lower.7ex\hbox{E}\kern-.125emX}}

\newcommand{\cK}{\mathcal{K}}

\newcommand{\cO}{\mathcal{O}}

\newcommand{\bh}{\mathbf{h}}

\newcommand{\bp}{\mathbf{p}}

\newcommand{\bs}{\mathbf{s}}

\newcommand{\bw}{\mathbf{w}}
\newcommand{\bx}{\mathbf{x}}

\newcommand{\bC}{\mathbf{C}}

\newcommand{\bH}{\mathbf{H}}
\newcommand{\bI}{\mathbf{I}}

\newcommand{\bQ}{\mathbf{Q}}

\newcommand{\bW}{\mathbf{W}}
\newcommand{\bX}{\mathbf{X}}

\newcommand{\bZ}{\mathbf{Z}}

\newcommand{\sinc}{\mathrm{sinc}}

\newcommand{\bbC}{\mathbb{C}}

\newcommand{\bLambda}{{\boldsymbol\Lambda}}

\newcommand{\bxi}{{\boldsymbol\xi}}
\newcommand{\bXi}{{\boldsymbol\Xi}}

\newcommand{\figref}[1]{Fig.~\ref{#1}}
\newcommand{\secref}[1]{Section~\ref{#1}}
\newcommand{\subsecref}[1]{Subsection~\ref{#1}}

\newcommand{\lemmref}[1]{\textit{Lemma}~\ref{#1}}
\newcommand{\tabref}[1]{Table~\ref{#1}}
\newcommand{\remref}[1]{\textbf{Remark~\ref{#1}}}

\newcommand{\trace}[1]{\mathrm{tr}\left(#1\right)}
\newcommand{\expect}[1]{\mathbb{E}{\left\{#1\right\}}}

\newcommand{\real}[1]{\Re\left\{#1\right\}}
\newcommand{\imag}[1]{\Im\{#1\}}

\newtheorem{lemma}{Lemma}
\newtheorem{remark}{Remark}

\begin{document}
\title{Spectral Efficiency Analysis of Near-Field Holographic MIMO over Ricean Fading Channels} 

\author{
	Mengyu~Qian, \IEEEmembership{Graduate Student Member,~IEEE,} 
	Xidong~Mu,~\IEEEmembership{Member,~IEEE,} 
	Li~You,~\IEEEmembership{Senior Member,~IEEE,}
Hyundong~Shin,~\IEEEmembership{Fellow,~IEEE,}
    and Michail~Matthaiou,~\IEEEmembership{Fellow,~IEEE}
	\thanks{Mengyu Qian and Li You are with the National Mobile Communications Research Laboratory, Southeast University, Nanjing 210096, China, and also with the Purple Mountain Laboratories, Nanjing 211100, China (e-mail: \{qianmy, lyou\}@seu.edu.cn).}
	\thanks{Xidong Mu is with the Centre for Wireless Innovation (CWI), Queen’s University Belfast, BT3 9DT Belfast, U.K. (e-mail:  x.mu@qub.ac.uk).}
\thanks{Hyundong Shin is affiliated with the Department of Electronic Engineering, Kyung Hee University, Yongin-si, Gyeonggi-do 17104, Republic of Korea (e-mail: hshin@khu.ac.kr).}
\thanks{Michail Matthaiou is with the Centre for Wireless Innovation (CWI), Queen’s University Belfast, BT3 9DT Belfast, U.K., and also affiliated with the Department of Electronic Engineering, Kyung Hee University, Yongin-si, Gyeonggi-do 17104, Republic of Korea (email: m.matthaiou@qub.ac.uk). The work of Michail Matthaiou has received funding from the European Research Council (ERC) under the European Union's Horizon 2020 research and innovation programme (grant agreement No. 101001331).}
\thanks{Part of this work was presented at the 2025 IEEE WCNC \cite{qianmengyu_HMIMO_conference}.}
}

\maketitle
\begin{abstract}
With the denser distribution of antenna elements, stronger mutual coupling effects would kick in among antenna elements, which would eventually affect the communication performance.  Meanwhile, as the holographic array usually has large physical size, the possibility of near-field communication increases. This paper investigates a near-field multi-user downlink HMIMO system and characterizes the spectral efficiency (SE) under the mutual coupling effect over Ricean fading channels. Both perfect and imperfect channel state information (CSI) scenarios are considered. (\romannumeral1) For the perfect CSI case, the mutual coupling and radiation efficiency model are first established. Then, the closed-form SE is derived under maximum ratio transmission (MRT). By comparing the SE between the cases with and without mutual coupling, it is unveiled that the system SE with mutual coupling might outperform that without mutual coupling in the low transmit power regime for a given aperture size. Moreover, it is also unveiled that the inter-user interference cannot be eliminated unless the physical size of the array increases to infinity. Fortunately, the additional distance term in the near-field channel can be exploited for the inter-user interference mitigation, especially for the worst case, where the users' angular positions overlap to a great extent. (\romannumeral2) For the imperfect CSI case, the channel estimation error is considered for the derivation of the closed-form SE under MRT. It shows that in the low transmit power regime, the system SE can be enhanced by increasing the pilot power and the antenna element density, the latter of which will lead to severe mutual coupling. In the high transmit power regime, increasing the pilot power has a limited effect on improving the system SE. However, increasing the antenna element density remains highly beneficial for enhancing the system SE. 
\end{abstract}


\section{Introduction}
The ongoing development and evolution of mobile wireless communication represents a paradigm shift in how we connect and communicate. Compared to 1G-5G, 6G is envisioned to support applications such as extended reality, autonomous systems, and massive machine-type communications. Hence, the demand for higher data rates, ultra-low latency, and unprecedented energy and spectral efficiency (SE) has become increasingly prominent \cite{theroadto6GMichail}. These requirements challenge conventional wireless technologies and drive the need for innovative solutions to accommodate the anticipated explosion of connected devices and immersive applications \cite{9144301}.

In line of this, holographic multiple-input multiple-output (HMIMO) is emerging as a transformative technology poised to meet the demands of the 6G era \cite{iacovelli2024holographic,10753352}. The key difference between HMIMO and massive MIMO is the antenna configuration. In HMIMO systems, both the antenna element size and inter-element spacing operate at sub-wavelength scale (typically much smaller than the wavelength), enabling ultra-compact implementations\cite{10232975}. In contrast, conventional massive MIMO arrays must maintain half-wavelength element spacing to minimize mutual coupling, resulting in substantially larger physical apertures when scaling up the antenna count. This fundamental distinction gives HMIMO its unique advantage in achieving extremely high integration densities while maintaining performance \cite{sanguinettiWavenumberDivisionMultiplexingLineofSight2023,wei2024electromagnetic}.

\begin{table*}[!t]
\centering
\ra{1.3}
\footnotesize
\caption{Our contributions in contrast to the state-of-the-art}
\begin{tabular}{LccccccccccR}
\toprule
& \cite{sanguinettiWavenumberDivisionMultiplexingLineofSight2023,gongHolographicMIMOCommunications2024} & \hspace{-0.08cm} \cite{weiTriPolarizedHolographicMIMO2023} & \hspace{-0.08cm} \cite{10262267} & \hspace{-0.08cm}\cite{dardariCommunicatingLargeIntelligent2020} & \hspace{-0.08cm} \cite{wangBeamformingPerformancesHolographic2024} & \hspace{-0.08cm}\cite{qian2024spectral}  & \hspace{-0.08cm}\cite{weiMultiUserHolographicMIMO2022} & \hspace{-0.08cm}\cite{liuDensifyingMIMOChannel2024} &\hspace{-0.08cm} \cite{10628002} &\hspace{-0.08cm} \cite{han2023superdirectivity} &\hspace{-0.08cm} \textbf{Proposed} \\ 
\midrule \rowcolor{lightblue}
Near field & \checkmark & \checkmark & \checkmark & \checkmark &$\times$ &\checkmark &\checkmark  &\checkmark &$\times$ &$\times$  &\checkmark \\ 
Mutual coupling  &$\times$  &$\times$ &$\times$ & $\times$ &\checkmark  &$\times$  & $\times$ &\checkmark &\checkmark &\checkmark &\checkmark\\  \rowcolor{lightblue}
Ricean fading channel  &$\times$  &$\times$  &$\times$  &$\times$  &$\times$  &$\times$  &$\times$ &$\times$ &$\times$ &$\times$ & \checkmark \\
Imperfect CSI &$\times$  &$\times$  &$\times$  &$\times$  &$\times$  &$\times$  &$\times$ &$\times$ &$\times$ &$\times$  & \checkmark \\ \rowcolor{lightblue}
Discrete antenna array & $\times$ &$\times$ &\checkmark  &$\times$   &\checkmark & $\times$ &$\times$ &\checkmark & \checkmark  & \checkmark & \checkmark \\ 
Multi-user &$\times$ &\checkmark &\checkmark &$\times$ &$\times$ &\checkmark &\checkmark  &\checkmark &\checkmark &\checkmark & \checkmark \\ \rowcolor{lightblue}
\bottomrule
\end{tabular}\label{Contrast_our_work}
\end{table*}

\subsection{Prior Works} 
Previous studies on HMIMO can generally be categorized into three aspects: research on channel modeling of HMIMO systems, analysis of how HMIMO enhances the system performance, and the design of transmission methods to achieve these performance improvements \cite{10232975}. We point out that the study of the new electromagnetic (EM) channel characteristics of HMIMO systems is necessary \cite{gongFieldChannelModeling2024,chenUnifiedFarFieldField2024,loykaInformationTheoryElectromagnetism2004}. Many existing works \cite{gongHolographicMIMOCommunications2024,jiangElectromagneticChannelModel2023,weiTriPolarizedHolographicMIMO2023,dardariHolographicCommunicationUsing2021} have combined EM field theory and utilized Green's functions to model channels considering only the line-of-sight (LoS) path. For example, the authors in \cite{gongHolographicMIMOCommunications2024} established the generalized EM-domain LoS channel models in a point-to-point near-field HMIMO systems with arbitrary surface placements, while the authors in \cite{10753352} obtained the channel coupling coefficients in a HMIMO communication system consisting of a transmitter and a receiver, both of which were equipped with holographic surfaces in a LoS scenario.

Meanwhile, acknowledging the necessity of non-line-of-sight (NLoS) paths in various scenarios, studies in \cite{liuDensifyingMIMOChannel2024,pizzoHolographicMIMOCommunications2020,weiMultiUserHolographicMIMO2022,pizzoFourierPlaneWaveSeries2022} have modeled HMIMO channels in complex scattering environments. For example,  the authors in  \cite{pizzoSpatiallyStationaryModelHolographic2020} provided a Fourier plane-wave series expansion of the channel response in a rich scattering environment.
Based on the EM channel model of HMIMO, researchers have analyzed the performance of HMIMO systems. One metric is the degrees of freedom (DoF) achievable by the HMIMO system \cite{10262267,dardariHolographicCommunicationUsing2021}. 
The authors in \cite{poonDegreesFreedomMultipleantenna2005} proposed a signal space approach for analyzing the DoF for far-field multi-antenna systems in scattering environments, suggesting that it is determined by the effective surface aperture and the angular spread. The authors in \cite{dardariCommunicatingLargeIntelligent2020} derived simple and analytical expressions for the link gain and the available spatial DoF in the system involving large intelligent surfaces, starting from EM arguments. 

At the same time, many studies have focused on analyzing the SE of HMIMO systems and designing effective holographic beamforming schemes, as HMIMO might achieve high beamforming gains and large spatial multiplexing \cite{jeonCapacityContinuousspaceElectromagnetic2018,wangBeamformingPerformancesHolographic2024,CanContinuousAperture,10628002,qian2024spectral}.  For example,  the theoretical expressions for the achievable SE performance in a multi-user HMIMO system with maximum-ratio-transmission (MRT) and zero-forcing (ZF) precoding schemes were derived in \cite{weiMultiUserHolographicMIMO2022}. The work was based on the multi-user EM-compliant channel models, which was expressed in the wavenumber domain using the Fourier plane wave approximation.
The capacity limits of continuous aperture array (CAPA)-based wireless communications were characterized in \cite{zhao2024continuous}, while the beamforming optimization in CAPA-based multi-user communications was studied in \cite{wang2024beamforming}. A concept called reconfigurable holographic surfaces was proposed to implement holographic radio and a holographic beamforming optimization algorithm was developed for beampattern gain maximization  in \cite{10163760}. Moreover, an EM hybrid beamforming scheme, based on a three-dimensional superdirective holographic antenna array, was proposed in \cite{10628002}. 
Besides, as the number of antenna elements in the array increases, the impact of hardware impairments (HWIs) on the system performance becomes increasingly severe. This is because deploying more antennas may lead to higher power consumption and more advanced circuitry, as well as unavoidable transceiver HWIs, such as the in-phase/quadrature-phase imbalance. In \cite{10480441}, the authors derived analytically  the SE and energy efficiency performance upper bound for reconfigurable holographic surfaces (RHS)-based beamforming architectures in the presence of HWIs, while in \cite{10502274}, the authors further considered both the phase shift error at the RHS elements and the HWIs at the radio frequeny chains of the transceivers and  proposed a hybrid beamforming architecture for near-field RHS-based cell-free networks.

\subsection{Motivations and Contributions}
Most existing studies on the SE of HMIMO systems overlook the effects of mutual coupling, since they often model the HMIMO array as a continuous surface \cite{sanguinettiWavenumberDivisionMultiplexingLineofSight2023,gongHolographicMIMOCommunications2024,weiTriPolarizedHolographicMIMO2023,weiMultiUserHolographicMIMO2022,dardariCommunicatingLargeIntelligent2020,qian2024spectral}. However, in practical implementations, discrete antenna arrays remain an essential configuration, which are composed of densely packed yet individually physical antennas.
In this case, a key issue in understanding the characteristics of HMIMO systems is how to intuitively grasp the effect of reduced array spacing and the resulting mutual coupling on the system performance. It is possible that mutual coupling may lead to a reduction in the radiated power \cite{yuanEffectsMutualCoupling2023}, but it could also enhance the system performance through super-directivity\cite{han2023superdirectivity,dovelos2023superdirective}.
To gain a comprehensive understanding of the operation of HMIMO systems, it is crucial to consider the mutual coupling effect inherent in discrete antenna arrays. The scientific community has attempted to reveal the impact of mutual coupling on the system SE through simulations in \cite{10628002,liuDensifyingMIMOChannel2024}. However, a theoretical analysis of the effect of mutual coupling on system SE is still missing.
Moreover, due to the typically large physical size of holographic arrays, the likelihood of communication occurring in the near field becomes significantly higher. Existing works \cite{sanguinettiWavenumberDivisionMultiplexingLineofSight2023,gongHolographicMIMOCommunications2024,weiTriPolarizedHolographicMIMO2023,weiMultiUserHolographicMIMO2022,dardariCommunicatingLargeIntelligent2020,qian2024spectral} do not adequately demonstrate the differences caused by the near-field and far-field channel model over the system SE.
Addressing these gaps can provide deeper insights into the interplay between densifying the antenna array and system's performance, offering valuable guidance for practical HMIMO system design. This is the main motivation of this work.

Against the above background, in this paper, we consider a near-field multi-user HMIMO downlink system for both perfect and imperfect channel state information (CSI) scenarios, assuming that the base station (BS) is equipped with a holographic planar array, while the users are all equipped with a single antenna.
For generality, we consider the holographic planar array as a discrete array and impose no constraints on the antenna spacing, thereby accounting for the mutual coupling effects inherently present in antenna arrays with any spacing (especially those smaller than half a wavelength). These effects on the transmitted signals and the overall system SE are analyzed theoretically and validated through simulations. The comparisons between our work and relevant state-of-the-art are listed in \tabref{Contrast_our_work}. 

The main contributions of our work are summarized as follows:

$\bullet$ We characterize the SE of a Ricean near-field muti-user HMIMO downlink system. Without imposing any restriction on the antenna element spacing of the holographic array at the BS, we first establish a mutual coupling model using an isotropic antenna model and a radiation efficiency model based on the Hannan's limit. Subsequently, we derive the expressions for the system SE with MRT under both perfect CSI and imperfect CSI conditions.

$\bullet$ For the perfect CSI case, we analyze the system SE with and without mutual coupling. We reveal that when the number of antenna elements tends to infinity, the corresponding SE is the upper bound of the system SE. However, when decreasing the antenna spacing of the array, we find that the SE with mutual coupling might outperform that without mutual coupling in the low transmit power regime.
Besides, we mathematically clarify  that the inter-user interference cannot be eliminated unless the physical size of the array increases to infinity. For practical limited aperture sizes, we find that the existence of the distance term in the near-field channel provides additional DoF to assist in distinguishing between different users, and thus, mitigate the inter-user interference.

$\bullet$ For the imperfect CSI case, we characterize the effects of both pilot power and mutual coupling on the system SE. In the low transmit power regime, the system SE can be enhanced by increasing the pilot power and the antenna element density, with the resulting mutual coupling further contributing to the SE improvement.
In the high transmit power regime, increasing the pilot power has a limited effect on improving the system SE,  but increasing the antenna element density remains highly beneficial for enhancing the system SE. However, such mitigation will be constrained by the radiation efficiency eventually.

$\bullet$ We validate the analytical results through simulation results. It is demonstrated that the SE of the HMIMO system is a trade-off between the radiation efficiency and the number of antenna elements. When the antenna radiation efficiency is acceptable, the system SE is always enhanced by increasing the antenna element density, which is particularly evident in the low transmit power regime. However, the antenna radiation efficiency may experience a fast reduction due to the existence of  mutual coupling as we densify the antenna array. 

\subsection{Organization and Notations}
The structure of the paper is as follows:
\secref{system_model} provides the system model for the near-field downlink multi-user HMIMO system with mutual coupling, including the channel model, transmit and receive signal models, and the SE model for each user.
\secref{asym_SE} presents an analysis of the system achievable SE under perfect and imperfect CSI,  with each scenario offering results from both perspectives: with and without mutual coupling.
\secref{num_res}  provides simulation results to validate the theoretical analyzes presented in the preceding sections.
\secref{conclusion} concludes the paper.

Throughout the paper, $\bX$ is a matrix, with $\bX^T$, $\bX^H$ being the transpose and conjugate transpose of $\bX$, respectively;
$\bx$ is a vector and $||\cdot||$ is the Euclidean norm; $\expect{\cdot}, \trace{\cdot}$ is the expectation operator and trace operator, respectively; $\mathcal{CN}(a,b)$ represents a Gaussian distribution, with $a$ being the mean and $b$ being the covariance; $\sinc(x) =\sin(\pi x)/(\pi x)$ is the normalized sinc function; $\real{\cdot}$ and $\imag{\cdot}$ represent the real and imaginary part of a complex value, respectively.

\begin{figure*}[t]
	\includegraphics[width=0.83\textwidth]{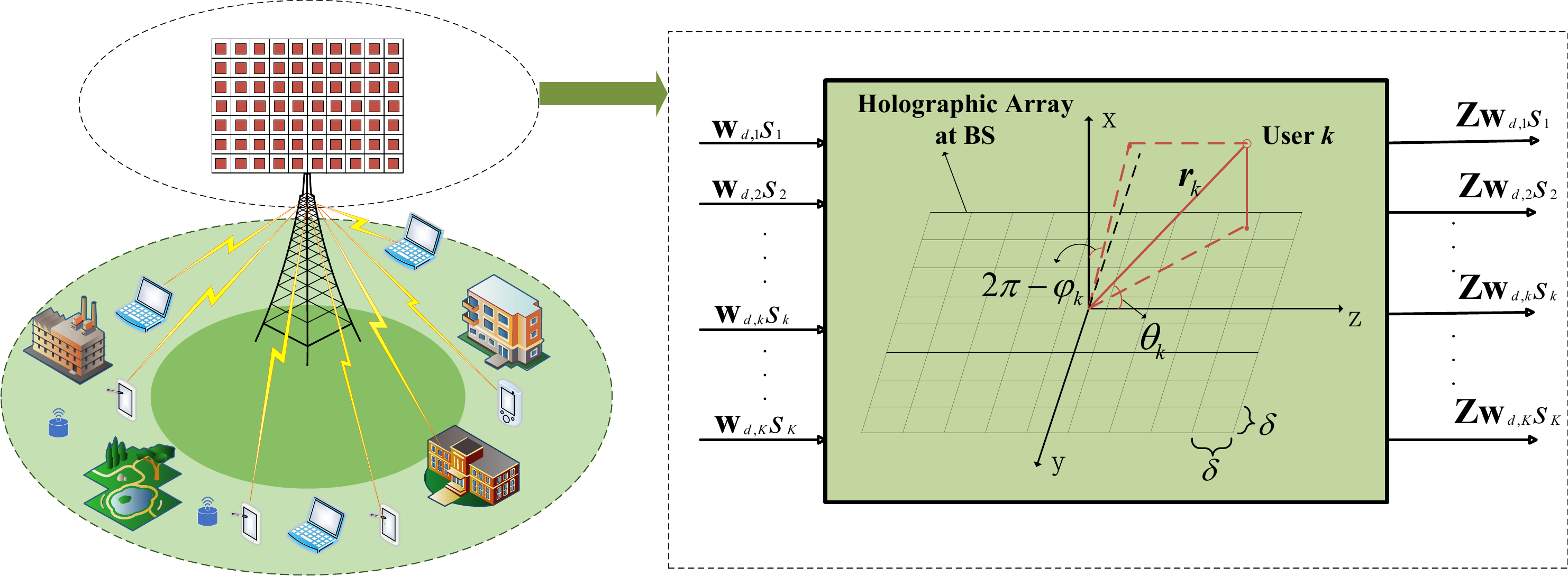}
	\centering
     \captionsetup{font={small}}
	\label{UE_array_position}
	\caption{Transmission between the BS and $K$ users.}
    \label{transmission_scenario}
\end{figure*}

\section{System Model}\label{system_model}
We consider a downlink HMIMO system where multiple single-antenna users are served by a BS equipped with a uniform planner array (UPA) having $N = N_y \times N_z$ antenna elements as shown in \figref{transmission_scenario}.   Without loss of generality, a three-dimensional (3D) Cartesian coordinate system is adopted.
The UPA lies in the $y$-$z$ plane and is centered at the origin, with $N_y$ and $N_z$ representing the number of antenna elements along the $y$-axis and $z$-axis, respectively.
For notational convenience, $N_y$ and $N_z$ are assumed to be odd numbers.
 
The position of the $\left(n_y, n_z\right)$th antenna element is denoted by $\mathbf{p}_{n_y, n_z}=\left[0, n_y \delta, n_z \delta\right]^T$, where $n_y=\left\{-\frac{N_y-1}{2},\cdots, -1,0, 1, \ldots,  \frac{N_y-1}{2}\right\}$ and $n_z=\left\{-\frac{N_z-1}{2},\cdots, -1,0, 1, \ldots,  \frac{N_z-1}{2}\right\}$. The physical dimensions of the UPA along the $y$-axis and $z$-axis are $N_y \delta$ and $N_z \delta$, respectively, where $\delta$ denotes the spacing between elements. 
As we focus on the radiated field density instead of the electric or magnetic field vectors, we assume that all antenna elements are isotropic with the same radiation power pattern and that the Maxwell equations still hold \cite{ivrlacCircuitTheoryCommunication2010}.

\subsection{Channel Model}\label{channel model}
HMIMO systems are typically suited for high frequencies and have large aperture sizes \cite{gongHolographicMIMOCommunications2024}, such that the near-field channel models have to be considered, where the generic spherical-wavefront model is employed to characterize the non-linear phase variations across different antenna elements. 
Hence, the $(n_y,n_z)$th element of the array response vector $\mathbf{a}\left(\theta_{k}, \varphi_{k}, r_{k}\right) \in \mathbb{C}^{N \times 1}$ is given by \cite{liuFieldCommunicationsTutorial2023}
\begin{align}\label{near-field array response vector}
	a^{\left(n_y, n_z\right)}\left(\theta_{k}, \varphi_{k}, r_{k}\right)=e^{-\jmath\frac{2\pi}{\lambda}r_k}\exp \left\{\!-\jmath \frac{2 \pi}{\lambda}\left(\!r_{k}^{\left(n_y, n_z\right)}-r_{k}\right)\!\right\},
\end{align}
where $\Delta r_{k}^{\left(n_y, n_z\right)} \triangleq r_{k}^{\left(n_y, n_z\right)}-r_{k}$ denotes the additional distance the wave travels to reach the $\left(n_y, n_z\right)$th antenna element relative to the $(0,0)$th element (i.e., reference element). It can be expressed as
\begin{align}\label{Delta_r_l}
&\Delta r_{k}^{\left(n_y, n_z\right)} \notag\\
&=r_{k}\left[\left(1+\left(\frac{r^{\left(n_y, n_z\right)}}{r_{k}}\right)^2-2 \frac{r^{\left(n_y, n_z\right)}}{r_{k}} g_{k}^{\left(n_y, n_z\right)}\right)^{1 / 2}-1\right],
\end{align}
where $r^{\left(n_y, n_z\right)} \triangleq \delta \sqrt{n_y^2+n_z^2}$ denotes the distance between the $\left(n_y, n_z\right)$th antenna element and the center one.
The geometric term $g_{k}^{\left(n_y, n_z\right)}$ is given by
\begin{subequations}
\begin{align}
 \!\!g_{k}^{\left(n_y, n_z\right)}\! &=\cos \vartheta^{\left(n_y, n_z\right)} \sin \theta_{k} \sin \varphi_{k}\!+\!\sin \vartheta^{\left(n_y, n_z\right)} \cos \theta_{k} \\
&=\frac{1}{\sqrt{n_y^2+n_z^2}}\left(n_y \sin \theta_{k} \sin \varphi_{k} + n_z \cos \theta_{k}\right),
\end{align}
\end{subequations}
with $\vartheta^{\left(n_y, n_z\right)}=\arctan \frac{n_y}{n_z}$ indicating the azimuth angle of any $\left(n_y, n_z\right)$th antenna element with respect to the reference one (i.e., $\left(n_y, n_z\right)=(0,0)$). 

By applying the Fresnel approximation \cite{liuFieldCommunicationsTutorial2023}, the additional distance in \eqref{Delta_r_l} can be re-expressed as 
\begin{subequations}
	\begin{align}
		& \Delta r_{k}^{\left(n_y, n_z\right)}   \approx \frac{1}{2}\frac{\left(r^{\left(n_y, n_z\right)}\right)^2}{r_{k}} - r^{\left(n_y, n_z\right)}  g_{k}^{(n_y, n_z)}
		\\& = \frac{\delta^2 \left(n_y^2+n_z^2\right) }{2 r_{k} } -  \delta \left(n_y \sin \theta_{k} \sin \varphi_{k} + n_z \cos \theta_{k} \right) \label{relative_phase}.
	\end{align}
\end{subequations}
Thus, the near-field Ricean fading channel $\bh_k$ between the BS and user $k$ can be modeled as
\begin{align}
  \bh_k = \sqrt{\frac{\mu_k}{1+\mu_k}} \beta_k \bh_{k, \rm LoS} + \sqrt{\frac{1}{1+\mu_k}} \tilde{\beta}_k \bh_{k, \rm NLoS},
\end{align}
where $\mu_k$ is the Ricean factor. The deterministic LoS channel is $\bh_{k, \rm LoS} = \mathbf{a}\left(\theta_{k}, \varphi_{k}, r_{k}\right)$, while $\bh_{k, \rm NLoS}$ denotes the random component. To more clearly analyze the impact of mutual coupling on the system SE, we have only considered the presence of mutual coupling as the main factor, while we assume that the entries of $\bh_{k, \rm NLoS}$ are independent and identically distributed (i.i.d) Gaussian random variables with zero mean and unit variance. Moreover,  $\beta_k$ and $\tilde{\beta}_k$ are the large-scale path loss coefficients. Since we consider the uniform spherical wave model in this paper, it is reasonable to assume that, $\forall n_y,n_z,$ $\beta_k^{(n_y,n_z)}  = \beta_k$ and $\tilde{\beta}_{k}^{(n_y,n_z)} = \tilde{\beta}_k$ \cite{liuFieldCommunicationsTutorial2023}.

\subsection{Ergodic SE under Mutual Coupling}
Let $\bs \in \bbC^{K\times 1} $ denote the signal vector with unit power that is from the BS, i.e.,  $\expect{\bs\bs^H} = \bI_K$, with $s_k$ being the $k$th element of $\bs$.
Denoting $\bW_d \in \bbC^{N\times K}$ as the digital beamforming matrix, the transmitted signal is
\begin{align}
	\bx = \bW_d\bs.
\end{align}

When the antenna elements on the transmit surface are placed in close proximity to each other,
the transmit signal will be altered because of the mutual coupling.
In this paper, the altered signal is denoted by $\tilde{\bx} = \bZ \bx$, with $\bZ \in \bbC^{N\times N}$ being the coupling transfer matrix, which is given by \cite{wangBeamformingPerformancesHolographic2024}
\begin{align}
	\bZ = \bC^{-1/2},
\end{align}
where $\bC = \{c_{n',n}\}\in \bbC^{N\times N}$ is the mutual coupling matrix, with $c_{n,n'}$ being the element in the $n$th column and $n'$th row of it. Here, $n = (n_z+\frac{N_z-1}{2})N_y + n_y + \frac{N_y-1}{2} +1 $.
By assuming that the radiation power pattern of each antenna element to be the same, $c_{n',n}$ can be further expressed as \cite{wangBeamformingPerformancesHolographic2024}
\begin{align}
  c_{n,n'} = \sinc{\left( \frac{2||\bp_n-\bp_{n'}||}{\lambda}\right) }.
\end{align}
As we consider lossy antennas, the mutual coupling matrix should be modified as $\tilde{\bC}$ \cite{ivrlacMultiportCommunicationTheory2014,han2023superdirectivity}, which is given by
\begin{align}
  \tilde{\bC} = \bC + \gamma \bI,
\end{align}
where $0\leq \gamma \leq 1$ is the loss factor. The presence of this loss factor arises from the fact that, in practical antenna systems, part of the energy is dissipated in the form of heat \cite{6880934,pizzoMutualCouplingHolographic2025}. When $\gamma = 0$, it means that there is no power dissipation, while $\gamma = 1$ means that the energy is all dissipated as heat.
Therefore, we have $\bZ = \tilde{\bC}^{-1/2}$.
The received signal of user $k$ is
\begin{align}
	y_k = \sqrt{\eta} & \bh_k^H \bZ \bw_{d,k} s_k  + \sqrt{\eta} \sum_{\substack{k'\in \cK \\ k'\neq k}} \bh_{k}^H \bZ  \bw_{d,k'} s_{k'} + n_k,
\end{align}
where $\cK \triangleq \{1,2,\dots,K\}$, while $\bw_{d,k}$ is the $k$th column of the digital beamforming matrix $\bW_{d} \triangleq [\bw_{d,1},\bw_{d,2},\dots,\bw_{d,K}]$. Moreover,  $n_k$ is the noise at the receiver with the distribution assumed to be $\mathcal{CN}(0,N_0)$ with $N_0$ being the power density of noise.
Due to the presence of mutual coupling, the transmitted power is not equal to the actual radiated power. Without loss of generality, we use the Hannan's limit to represent the radiation efficiency of each antenna element in the following, which is given by \cite{yuanEffectsMutualCoupling2023,liElectromagneticInformationTheory2023}
\begin{align}
  \eta = \frac{\pi d_x d_y}{\lambda^2}.
\end{align}
Here, $d_x$ and $d_y$ are both assumed to be equal to $\delta$.  Note that the presence of mutual coupling may reduce the radiation efficiency and, thus,  leads to a lower radiated power. If mutual coupling is not considered, $\eta$ is assumed to be $1$ and $\bZ = \bI$.

Therefore, the ergodic SE (bps/Hz) of user $k$ under the existence of mutual coupling is given by
\begin{align}\label{sum_rate_k}
	R_{k}\left(\bW_d \right) =  \expect{\log_2 \left( 1 +\gamma_k\left(\bW_d\right) \right)},
\end{align}
where
\begin{align}\label{SINR}
	\gamma_k(\bW_d) =  \frac{ \eta \left|\bh_k^H \bZ \bw_{d,k}\right|^2}{\eta \sum_{\substack{k'\in \cK \\ k'\neq k}}  \left| \bh_{k}^H \bZ  \bw_{d,k'} \right|^2  + N_0}.
\end{align}
Considering that the system SE is simply the sum of the individual SEs of each user, i.e., $R = \sum_{k=1}^{K}R_k(\bW_d)$, in the subsequent analysis, we will focus on the analysis of $R_k(\bW_d)$ without loss of generality.

\section{Analysis of Achievable SE}\label{asym_SE}
In this section, we will first consider the achievable SE under perfect CSI in two cases: with and without mutual coupling. After comparing the SE in these two cases, we will provide some useful design insights for the pure LoS and Rayleigh fading scenarios, which represent the two extreme cases of the Ricean fading channel.
Following this, we will then extend the perfect CSI analysis to the case of imperfect CSI, and similarly analyze the achievable SE with and without mutual coupling as well as the design insights.
\subsection{Perfect CSI}
Here, MRT is assumed to be employed at the BS for each user with perfect CSI, which is given by
\begin{align}\label{MRT_precoding}
	\bW_d = \alpha_{\rm MRT} \bH^H,
\end{align}
where $\bH \triangleq [\bh_{1}^H,\bh_{2}^H,\dots,\bh_{K}^H]^H$. We assume that the power is equally distributed among the users, while $\alpha_{\rm MRT}$ is a normalization coefficient to ensure that the average transmit power constraint is satisfied, i.e.,  
\begin{align}
	\expect{ \trace{ \bW_d   \bW_d^H } } = p_u.
\end{align}
Using \eqref{MRT_precoding}, the SINR in \eqref{SINR} becomes
\begin{align}
	\gamma_k(\bW_d) = \frac{\alpha_{\rm MRT}^2 \eta \left|\bh_k^H \bZ \bh_k \right|^2}{N_0 + \alpha_{\rm MRT}^2 \eta \sum_{\substack{k'\in \cK \\ k'\neq k}}  \left| \bh_{k}^H \bZ \bh_{k'} \right|^2 }.
\end{align}

As isotropic antennas are considered, the coupling transferring matrix $\bZ = \tilde{\bC}^{-1/2}$ is a real symmetric matrix that can be decomposed as $\bZ = \bQ \bLambda \bQ^T$. 
As the array becomes denser, additional eigenvectors associated with extremely small eigenvalues are introduced into the eigenspace of the mutual coupling matrix $\tilde{\bC}$. Consequently, this leads to the emergence of corresponding eigenvectors with very large eigenvalues, which are approximately the reciprocal halves of the small eigenvalues. Thus, $\bLambda$ is a diagonal matrix with the each entry being positive and larger than $1$ \cite{wangBeamformingPerformancesHolographic2024}.

The ergodic SE of user $k$ is
\begin{align}\label{expectations_SE}
  R_k = \expect{\log_2\left(1+ \frac{\alpha_{\rm MRT}^2 \eta ||\tilde{\bh}_k ||^4}{ N_0 + \alpha_{\rm MRT}^2 \eta \sum_{\substack{k'\in \cK \\ k'\neq k}}  \left| \tilde{\bh}_k^H \tilde{\bh}_{k'} \right|^2 }\right)},
\end{align}
where $\tilde{\bh}_k = \bLambda^{\frac{1}{2}} \bQ^T \bh_k $ is the equivalent channel with mutual coupling.
The SE without mutual coupling is a special case of that with mutual coupling. Thus, we derive the achievable SE with mutual coupling and then obtain the achievable SE without mutual coupling by setting $\bZ = \bI$.
\subsubsection{Achievable SE with Mutual Coupling}
Let $\bh_k$ (or $\bH$) denote $\bQ^T \bh_k $ (or $\bQ^T\bH$) since each pair is unitary equivalent. Then, $\tilde{\bh}_k$ can be expressed as $\tilde{\bh}_k = \bLambda^{\frac{1}{2}} \bh_k $, and by definition, we have
\begin{align}\label{expect_hk_hi_withMC}
\expect{\tilde{\bh}_k^H \tilde{\bh}_i} 
 \triangleq \frac{\sqrt{\mu_k \mu_i}}{ \sqrt{\left(\mu_k+1\right)\left(\mu_i+1\right)}} \tilde{Q}(k,i),
\end{align}
where
\begin{align}
  \tilde{Q}(k,i)& \triangleq \sum_{n_y} \sum_{n_z}\lambda_{(n_y,n_z)} f_{n_y}(k,i)   f_{n_z}(k,i),
\end{align}
and
\begin{subequations}
  \begin{align}
    & f_{n_y}(k,i)\triangleq e^{\jmath \frac{2\pi}{\lambda} \left[  \delta^2 n_y^2 \left( \frac{1}{2 r_{k} } -  \frac{1}{2 r_{i}} \right)  - \delta  n_y \left( \sin \theta_{k} \sin \varphi_{k} - \sin \theta_{i} \sin \varphi_{i} \right) \right] }, \\
   & f_{n_z}(k,i) \triangleq  e^{\jmath \frac{2\pi}{\lambda} \left[  \delta^2 n_z^2 \left( \frac{1}{2 r_{k} } -  \frac{1}{2 r_{i}} \right)  - \delta  n_z \left( \cos \theta_{k}  - \cos \theta_{i}  \right) \right] }.
  \end{align}
\end{subequations}

To obtain the achievable SE with mutual coupling, we first introduce the lemma below.
\begin{lemma}\label{P_MC_hk_hi_square_lemma}
    The expectation of the inner product of two same columns in $\tilde{\bH}$ is given by
  \begin{align}
    \expect{||\tilde{\bh}_k||^2} = \expect{\tilde{\bh}_k^H \tilde{\bh}_k} = \trace{\bLambda},
  \end{align}
  and the expectation of the norm square of the inner product of any two columns in $\tilde{\bH}$ is given by
  \begin{align}
    \expect{|\tilde{\bh}_k^H \tilde{\bh}_i|^2} = 
    \begin{cases}
   \trace{\bLambda}^2+ \frac{(2\mu_k+1)\trace{\bLambda^2}}{(1+\mu_k)^2}, & k=i, \\ 
    \frac{\mu_k\mu_i  \big|\tilde{Q}(k,i)\big|^2  + \trace{\bLambda^2}(\mu_k+\mu_i+1) }{(\mu_k+1)(\mu_i+1)}, & k \neq i.
    \end{cases}
  \end{align}
\end{lemma}

\begin{proof}
See Appendix \ref{hk_hi_norm_square_lemma}.
\end{proof}

Using \lemmref{P_MC_hk_hi_square_lemma} and \cite[Lemma 1]{zhangPowerScalingUplink2014}, we have the approximated SE of user $k$ in \eqref{rate_k_MRT_Ricean} at the top of next page,
\begin{figure*}[ht]
\begin{align}\label{rate_k_MRT_Ricean}
  R_{\rm MC}^{P,k} \approx \tilde{R}_{\rm MC}^{P,k} = \log_2\left( 1+  \frac{\alpha^2_{MRT} \eta \left[ (1+\mu_k)^2 \trace{\bLambda}^2 + (2\mu_k+1)\trace{\bLambda^2}\right]}{N_0(1+\mu_k)^2 + \alpha^2_{MRT} \eta (1+\mu_k)\sum_{k' \neq k} \tilde{\Delta}_{kk'}} \right).
\end{align}
\hrulefill
\end{figure*}
where
\begin{align}
  \tilde{\Delta}_{k k'} = \frac{\mu_k\mu_{k'}  \big|\tilde{Q}(k,k')\big|^2  + \trace{\bLambda^2}(\mu_k+\mu_{k'}+1)}{\mu_{k'}+1}.
\end{align}

\subsubsection{Achievable SE without Mutual Coupling} By substituting $\bZ = \bI$ and $\eta = 1$, we can similarly obtain the approximated SE in the case of non-mutual coupling, which is given by
\begin{align}\label{rate_nMC_k_MRT_Ricean}
   & R_{nMC}^{P,k}  \approx \tilde{R}_{nMC}^{P,k} \notag \\
   & = \log_2\left(1+ \frac{\alpha_{\rm MRT}^2 \eta \expect{||\tilde{\bh}_k ||^4}}{ N_0 + \alpha_{\rm MRT}^2 \eta \sum_{\substack{k'\in \cK \\ k'\neq k}} \expect{ \left| \tilde{\bh}_k^H \tilde{\bh}_{k'} \right|^2 }}\right),\notag\\
   & = \log_2\left(1+ \frac{\alpha_{\rm MRT}^2 \left[(2\mu_k + 1)N + N^2(\mu_k+1)^2\right]}{\alpha_{\rm MRT}^2 (\mu_k+1)\sum_{k' \neq k} \Delta_{k k'} + N_0(\mu_k+1)^2}\! \!\right),
\end{align}
with
\begin{align}\label{Delta_kk'}
  \Delta_{k k'} = \frac{\mu_k \mu_{k'} \left|Q(k,k')\right|^2  + N(\mu_k + \mu_{k'}+1)}{1+\mu_{k'}}.
\end{align}
The term $\left|Q(k,k')\right|^2$ in \eqref{Delta_kk'} represents the inter-user interference, which is given by
\begin{align}\label{Q_ki}
   Q\left(k,i\right) \! \! \triangleq \! \!& \int_{-\frac{N_y}{2}-\frac{B_{k,i}}{2A_{k,i}}}^{\frac{N_y}{2}-\frac{B_{k,i}}{2A_{k,i}}} e^{\jmath \pi \cdot A_{k,i} x^2 }  d x \cdot \! \! \! \int_{-\frac{N_z}{2}-\frac{D_{k,i}}{2A_{k,i}}}^{\frac{N_z}{2}-\frac{D_{k,i}}{2A_{k,i}}} e^{\jmath \pi \cdot A_{k,i} x^2 },
\end{align}
and
\begin{subequations}\label{parameters_ABD}
  \begin{align}
  A_{k,i} =& \frac{\delta^2 }{\lambda} \left(\frac{1}{r_k}- \frac{1}{r_i}\right), \\
  B_{k,i} =& \frac{2\delta}{\lambda} \left(\sin \theta_{k} \sin \varphi_{k} - \sin \theta_{i} \sin \varphi_{i}\right),\\
  D_{k,i} =& \frac{2\delta}{\lambda} \left(\cos \theta_{k} - \cos \theta_{i} \right).
  \end{align}
\end{subequations}

The expressions of \eqref{rate_k_MRT_Ricean} and \eqref{rate_nMC_k_MRT_Ricean} have similar structures, except that the position of the number of antennas $N$ in \eqref{rate_nMC_k_MRT_Ricean} is replaced by a function of the coupling matrix $\trace{\bLambda}$ in \eqref{rate_k_MRT_Ricean}. This observation implies that the mutual coupling present in the antenna array can be regarded as a factor that affects system performance, similar to the channel correlation. However, mutual coupling is an inherent property of the antenna array and is not influenced by the scattering environment. Note that the expressions in \eqref{rate_k_MRT_Ricean} and \eqref{rate_nMC_k_MRT_Ricean} involve multiple factors, including the mutual coupling matrix, inter-user interference, and the Ricean factor, which render the performance analysis quite challenging. Therefore, to provide more design insights, we will analyze two special cases in the following.

\subsubsection{Achievable SE under Pure LoS and Rayleigh Fading Scenarios}
When $\mu_k = 0$, the primary source of interference in the system originates from background noise, which can be observed from \eqref{Ray_rate_MC_P} and \eqref{Ray_rate_nMC_P} in the following. However, as $\mu_k$ increases, the proportion of inter-user interference $\big|Q(k,k')\big|^2$ (or $|\tilde{Q}(k,k')|^2$ in the mutual coupling case) in the system rises. At this stage, it becomes necessary to comprehensively consider the impact of both background noise and inter-user interference on the system SE. To better illustrate this, the subsequent analysis will consider two extreme cases of Ricean fading, namely pure LoS and Rayleigh fading scenarios, to demonstrate the impact of mutual coupling on the system SE under different dominant interference sources.  
For each scenario, we will analyze the variation of SE with an increasing number of array elements, and discuss the design insights.
\paragraph{Rayleigh Fading Scenario}
When $\mu_k = 0, \forall k \in \cK$,  \eqref{rate_k_MRT_Ricean} and \eqref{rate_nMC_k_MRT_Ricean} reduce respectively to the following expressions,
\begin{align}\label{Ray_rate_MC_P}
\tilde{R}_{\rm MC,Ray}^{P,k}
 &= \log_2\left( 1+  \frac{\alpha^2_{MRT} \eta \left[ \trace{\bLambda}^2 / \trace{\bLambda^2} + 1 \right]}{N_0/ \trace{\bLambda^2} + \alpha^2_{MRT} \eta (K-1) } \right),
\end{align}
and
\begin{align}\label{Ray_rate_nMC_P}
  \tilde{R}_{\rm nMC,Ray}^{P,k} = \log_2\left(1+ \frac{\alpha_{\rm MRT}^2 \left(N^2+ N \right)}{\alpha_{\rm MRT}^2 (K-1)N + N_0}\right).
\end{align}
In this case,  the only interference comes from the term of $N_0$, which will negatively affect the system SE, especially in the low transmit power regime.

\begin{figure}[htbp]
    \captionsetup{font={small}}
    \centering 
    \includegraphics[width = 0.45\textwidth]{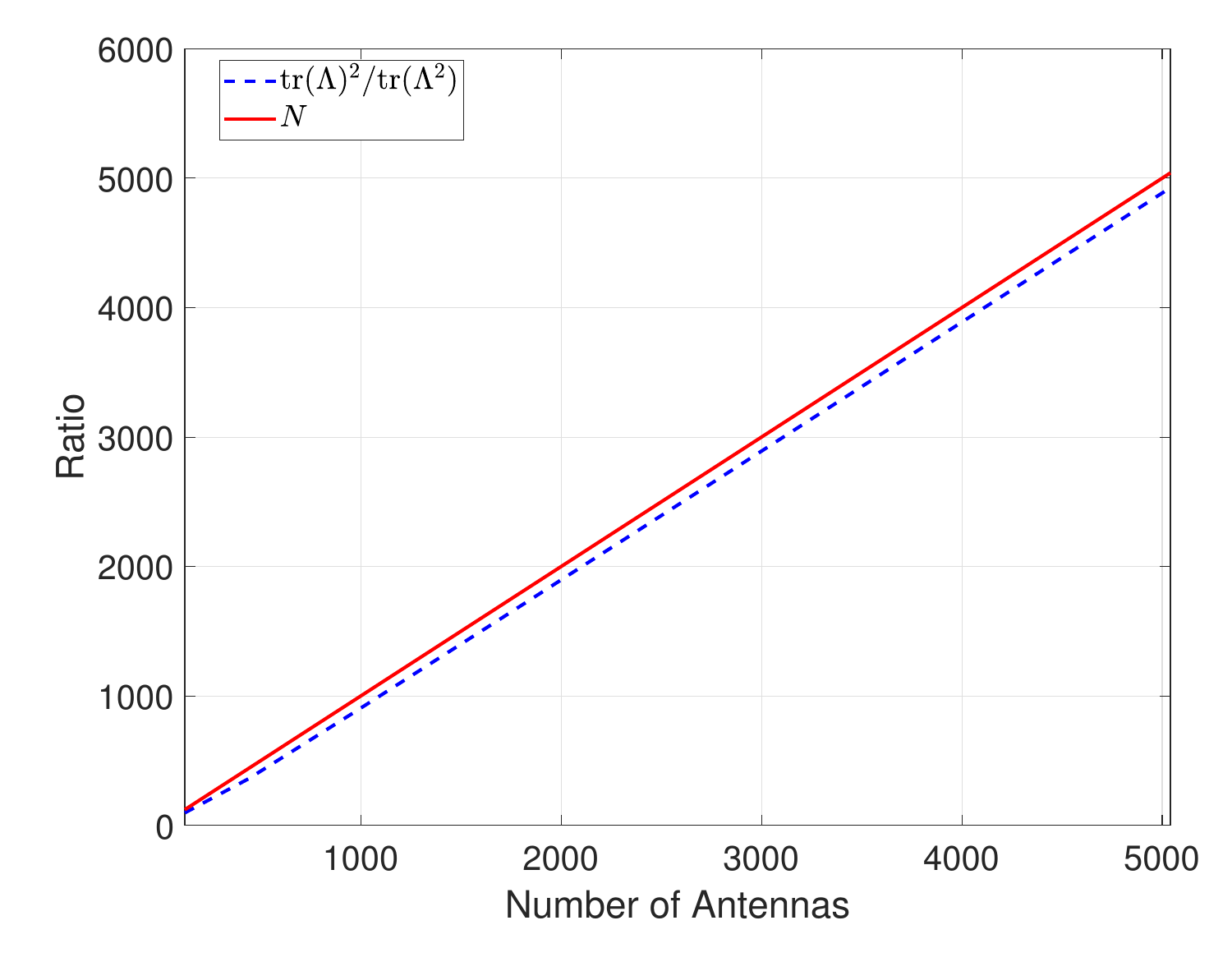}
     \caption{Ratio versus the number of antennas.}
     \label{fig_ratio}
\end{figure}

For \eqref{Ray_rate_MC_P},  when the antenna array is dense enough, i.e., $N \rightarrow \infty$, both $\trace{\bLambda}^2$ and $\trace{\bLambda^2}$ are likely to approach infinity since each element of $\bLambda$ is quite larger than $1$ \cite{wangBeamformingPerformancesHolographic2024}.  To better illuminate the ratio of $\trace{\bLambda}^2/\trace{\bLambda^2}$, let $\Lambda_{i} = \alpha_i \trace{\bLambda}$ denote the element of $\bLambda$, where $0 \leq \alpha_i \leq 1$ and $\sum_{i=1}^{N} \alpha_i = 1$. 
When $N$ increases, the ratio of $\trace{\bLambda}^2/\trace{\bLambda^2} $ can be approximated as
\begin{align}\label{ratio_lambda}
  \frac{\trace{\bLambda}^2}{\trace{\bLambda^2}} = \frac{1}{\sum_{i = 1}^{N}\alpha_i^2} \approx N,
\end{align}
which is also verified in \figref{fig_ratio}.

Therefore, \eqref{Ray_rate_MC_P} can be approximated by 
\begin{align}
  \tilde{R}_{\rm MC,Ray}^{P,k} \approx & \log_2\left( 1+  \frac{\alpha^2_{MRT} ( N + 1 )}{N_0/ [\eta\trace{\bLambda^2}] + \alpha^2_{MRT} (K-1) }\right).
\end{align}
Substituting the Hannan's limit into $\eta \trace{\bLambda^2}$, we have 
\begin{align}
  \eta \trace{\bLambda^2} = \frac{\pi L_x L_y}{\lambda^2}\frac{\trace{\bLambda^2}}{N}.
\end{align}
As $\trace{\bLambda^2} \sim \cO(N)$, $\trace{\bLambda^2}/N$ converges to a constant in the extreme case of $N \rightarrow \infty$, yielding
\begin{align}
  \frac{\pi L_x L_y}{\lambda^2}\frac{\trace{\bLambda^2}}{N} \propto \frac{\pi L_x L_y}{\lambda^2}.
\end{align}
\begin{remark}\label{remark_ray}
In the low transmit power regime, where $N_0$ cannot be ignored, $\tilde{R}_{\rm MC,Ray}^{P,k}$ may outperform $\tilde{R}_{\rm nMC,Ray}^{P,k}$.
This is because $\trace{\bLambda^2}$ is larger than $N$, making it possible that $\eta\trace{\bLambda^2} \geq N$ when the radiation efficiency is acceptable. However, when $N$ increases to a certain extent, the system's radiation efficiency $\eta$ becomes very low. In the extreme case of $N \rightarrow \infty$,
$\tilde{R}_{\rm nMC,Ray}^{P,k}$ serves as the upper bound of the system SE, since the term of  $N_0/ \left(\eta\trace{\bLambda^2}\right)$ will not diminish to zero, while the noise $N_0$ will be eliminated in \eqref{Ray_rate_nMC_P}.
At this point, the system SE without mutual coupling will undoubtedly outperform that with mutual coupling.
In other words, the SE of the system is relevant to the trade-off between the radiation efficiency $\eta$ and the number of antenna elements.  In the high transmit power regime, the impact of $N_0$ can be neglected. Under this condition, the system SE with and without considering mutual coupling will exhibit minimal difference, with the former being slightly lower than the latter.
\end{remark}

\paragraph{LoS Scenario}
When $\mu_k \rightarrow \infty, \forall k \in \cK$, i.e., the LoS-dominant scenario, the approximations in \eqref{rate_k_MRT_Ricean} and \eqref{rate_nMC_k_MRT_Ricean} become
\begin{align}\label{LoS_rate_MC_P}
& \tilde{R}_{\rm MC,LoS}^{P,k} \notag \\
& \rightarrow \log_2\left( 1+  \frac{\alpha^2_{MRT} \eta \trace{\bLambda}^2}{N_0 + \alpha^2_{MRT}  \eta \sum_{k' \neq k} \big|\tilde{Q}(k,k')\big|^2}\right),
\end{align}
and
\begin{align}\label{LoS_rate_nMC_P}
	& \tilde{R}_{\rm nMC,LoS}^{P,k} \rightarrow  \log_2 \left( \! 1+ \frac{\alpha_{\rm MRT}^2  N^2}{\alpha_{\rm MRT}^2 \sum_{k' \neq k} \left|Q(k,k')\right|^2 + N_0} \!\right)\!.
\end{align}
With MRT, the LoS rate is largely constrained by the inter-user interference $Q(k,k')$ and $\tilde{Q}(k,k')$, which will be discussed in the following.

$\bullet$ Values of $|Q(k,k')/N|^2$: When $N$ increases, the ratio of $Q(k,k')/N$ will change and it is the only source of interference in \eqref{LoS_rate_nMC_P}. Thus, we choose to analyze it instead of $Q(k,k')$.

As part of $Q(k,k')/N$, define
\begin{align}
 \frac{1}{N_y}  \int_{-\frac{N_y}{2}-\frac{B_{k,i}}{2A_{k,i}}}^{\frac{N_y}{2}-\frac{B_{k,i}}{2A_{k,i}}} e^{\jmath \pi \cdot A_{k,i} x^2 } d x
 \triangleq F_1 + F_2,
\end{align}
where
\begin{subequations}\label{F_1}
  \begin{align}
    F_1 &= \frac{1}{N_y}  \int_{0}^{\frac{N_y}{2}-\frac{B_{k,i}}{2A_{k,i}}} e^{\jmath \pi \cdot A_{k,i} x^2 } d x \\
    &\overset{(a)}{=} \frac{1}{L_y} \int_{0}^{c_1(L_y)} e^{\jmath \pi \cdot \frac{1}{\lambda}(\frac{1}{r_k}- \frac{1}{r_i}) t^2 } d t.
  \end{align}
\end{subequations}
Here, $(a)$ represents the replacement of $t \triangleq \delta x$; $L_y$ is the length of the transmit holographic array along the $y$ axis while $c_1(L_y)$ is given by
\begin{align}
  c_1(L_y) = \frac{L_y}{2} - \frac{\sin \theta_{k} \sin \varphi_{k} - \sin \theta_{i} \sin \varphi_{i}}{\frac{1}{r_k}- \frac{1}{r_i}}.
\end{align}
Note that $F_2$ can be similarly obtained, yielding
\begin{subequations}\label{F_2}
\begin{align}
  F_2 &= \frac{1}{N_y}  \int_{0}^{\frac{N_y}{2}+\frac{B_{k,i}}{2A_{k,i}}} e^{\jmath \pi \cdot A_{k,i} x^2 } d x \\
    &\overset{(a)}{=} \frac{1}{L_y} \int_{0}^{c_2(L_y)} e^{\jmath \pi \cdot \frac{1}{\lambda}(\frac{1}{r_k}- \frac{1}{r_i}) t^2 } d t,
\end{align}
\end{subequations}
where
\begin{align}
   c_2(L_y) = \frac{L_y}{2} + \frac{\sin \theta_{k} \sin \varphi_{k} - \sin \theta_{i} \sin \varphi_{i}}{\frac{1}{r_k}- \frac{1}{r_i}}.
\end{align}
Referring to the properties of Fresnel functions \cite{Fresnel_ref}, the integrals in $F_1$ and $F_2$ are always both finite, and this, leads to the finite values of $|Q(k,k')/N|^2$ when $N \rightarrow \infty$.
Reducing the effects of inter-user interference $|Q(k,k')/N|^2$ in \eqref{LoS_rate_nMC_P} is only possible if $L_y$ and $L_z$ are increasing to infinity.

$\bullet$ Values of $|\tilde{Q}(k,i)|^2$:  The specific elements of $|\tilde{Q}(k,i)|^2$ are
\begin{align}\label{tilde_Q_specific}
  |\tilde{Q}(k,i)|^2 = \sum_{n = 1}^{N}\lambda_n^2 + 
 \sum_{\substack{p,q  = 1,\\ p \neq q}}^{N} 2\lambda_p \lambda_q \cos(x_p - x_q),
\end{align}
where $p \triangleq {(n_z+\frac{N_z-1}{2})N_y+n_y+\frac{N_y-1}{2} +1}$ and 
\begin{align}\label{element_tilde_Q}
  & x_{p} = \frac{2\pi}{\lambda}\left[\delta^2( n_y^2 + n_z^2)(\frac{1}{2 r_{k} } -  \frac{1}{2 r_{i}}) - \right. \notag \\
  & \!\left. \delta \left[n_y \left( \sin \theta_{k} \sin \varphi_{k} - \sin \theta_{i} \sin \varphi_{i} \right) \! + \!  n_z \left( \cos \theta_{k}  - \cos \theta_{i}  \right) \right] \right].
\end{align}
The term $\sum_{k' \neq k} \big|\tilde{Q}(k,k')\big|^2$ is the inter-user interference, and it is mostly affected by the users' positions and the physical size of the antenna.  In other words, when the users are randomly distributed within the system, it is difficult to determine whether inter-user interference is greater or smaller when mutual coupling is taken into account.
In the worst case, the inter-user interference $|\tilde{\bQ}(k,i)|^2 = \trace{\bLambda}^2$ and $|\bQ(k,i)|^2 = N^2$, which is possible when the users are distributed densely and the phase differences between different users are too small to be distinguished. In this scenario, it is necessary to drop some users from service \cite{6951994}.   
\begin{remark}\label{near_far_remark}
Unlike the far-field, where only linear terms of elevation angles and azimuth angles exist, the near-field channel model incorporates differences in the distance dimension, which can reduce the likelihood of inter-user interference approaching the worst-case scenario. Even if two users are located at the same elevation and azimuth angles, as long as there is a difference in their distances, the system performance will not reach the aforementioned worst-case scenario.  Although the complete elimination of inter-user interference requires the array physical size to approach infinity, the additional spatial dimension introduced by the near-field channel helps mitigate the impact of inter-user interference on the system SE.
\end{remark}

To better illuminate the performance between $\tilde{R}_{nMC,LoS}^{P,k}$ and $\tilde{R}_{MC,LoS}^{P,k}$, we look into the system SE in the worst case:
\begin{subequations}
  \begin{align}
    \tilde{R}_{\rm MC,LoS}^{P,k} &= \log_2\left( 1+  \frac{\alpha^2_{MRT} }{N_0/[\eta \trace{\bLambda}^2] + \alpha^2_{MRT} (K-1)}\right),\\
    \tilde{R}_{\rm nMC,LoS}^{P,k} &= \log_2\left( 1+  \frac{\alpha^2_{MRT} }{N_0/N^2 + \alpha^2_{MRT} (K-1)}\right) \label{LoS_infty}.
  \end{align}
\end{subequations}
As $\trace{\bLambda}^2$ is larger than $N^2$, it is possible that $\eta \trace{\bLambda}^2 \geq N^2$, thus, leading to $\tilde{R}_{\rm MC,LoS}^{P,k} \geq \tilde{R}_{\rm nMC,LoS}^{P,k}$ for low transmit power.

On the other hand, when $N \rightarrow \infty$, $\trace{\bLambda}^2 \sim \cO(N^2)$ and
\begin{align}
  \eta \trace{\bLambda}^2 = \frac{\pi L_x L_y}{\lambda^2}\frac{\trace{\bLambda}^2}{N} \rightarrow \cO(N).
\end{align}
Thus, $\tilde{R}_{\rm MC,LoS}^{P,k}$ will converge to $\tilde{R}_{\rm nMC,LoS}^{P,k}$ as $N_0$ in \eqref{LoS_infty} will be eliminated more emphatically.

\begin{remark}\label{remark_los}
By comparing the system' SE with and without mutual coupling in the worst case, we can draw similar conclusions: under a low transmit power budget, where the background noise $N_0$ cannot be ignored compared with the transmitted signal power,
$\tilde{R}_{\rm MC,LoS}^{P,k}$ might also outperform $\tilde{R}_{\rm nMC,LoS}^{P,k}$ as long as the radiation efficiency $\eta$ is acceptable. However, in practical cases,  the radiation efficiency $\eta$ will experience a fast reduction when we densify the antenna array. Thus, when the antenna spacing is too small, $\tilde{R}_{\rm MC,LoS}^{P,k}$ tends to be smaller than $\tilde{R}_{\rm nMC,LoS}^{P,k}$. 
In the extreme case of $N \rightarrow \infty$, $\tilde{R}_{\rm MC,LoS}^{P,k}$ will converge to $\tilde{R}_{\rm nMC,LoS}^{P,k}$ finally, as the noise term will converge to zero more quickly in $\tilde{R}_{\rm nMC,LoS}^{P,k}$.
In the high transmit power regime,  $\tilde{R}_{\rm MC,LoS}^{P,k}$ and $\tilde{R}_{\rm nMC,LoS}^{P,k}$ exhibit almost identical performance.
\end{remark}

From the above theoretical analysis, we can conclude the following::
1) The SE with a continuous antenna array ($N \rightarrow \infty$) is the upper bound of the discrete case; 2) When $N$ is moderate and $\eta$ is acceptable, the SE with mutual coupling is possible to outperform that without mutual coupling; 3) The inter-user interference will be eliminated when the physical size of the antenna array is infinitely large; and 4) The near-field will prevent the system SE from being the worst case since it offers the extra distance dimension to assist in distinguishing different users. 
All the conclusions mentioned above will also be validated in the subsequent simulations.

\subsection{Imperfect CSI}\label{subsec_IP}
In practice, the channel matrix $\bH$ is estimated at the BS. Assume that the  length of the pilot symbols is $p$ ($p \geq K$). Then, the power of the pilots is $\tau \triangleq p p_u/K$.
Let $\bXi = \hat{\bH}-\bH \triangleq \left[\bxi_1,\bxi_2,\dots,\bxi_K\right]$ denote the channel estimation error.
The variance of the elements of the estimation error vector $\bxi_k$ is \cite{zhangPowerScalingUplink2014}
\begin{align}
  \expect{ \big| [\bXi]_{n,k}-\expect{[\bXi]_{n,k}} \big|^2} = \frac{1}{(1+\tau)(1+\mu_k)}.
\end{align}

In this case, the received signal of the $k$th user is
\begin{align}
  \hat{y}_k =& \sqrt{\eta} \hat{\bh}_k^H \bZ \hat{\bw}_k s_k + \sqrt{\eta} \sum_{\substack{k'\in \cK \\ k'\neq k}}\hat{\bh}_k^H \bZ \hat{\bw}_{k'} s_{k'} \notag \\
  &- \sqrt{\eta} \sum_{i=1}^{K} \bxi_k^H \bZ \hat{\bw}_i s_{i} + n_k.
\end{align}
Let $\hat{\bW}_d \triangleq [\hat{\bw}_1,\hat{\bw}_2,\dots,\hat{\bw}_K]$ denote the MRT beamforming matrix which depends on $\hat{\bH}$, i.e., $\hat{\bW}_d \triangleq \alpha_{\rm MRT} \hat{\bH}$.   As a result, the ergodic SE of the $k$th user can be written as \eqref{achievable_k_rate} shown at the top of next page.
\begin{figure*}[t]
\begin{align}\label{achievable_k_rate}
  R_{IP,k} = \expect{ \log_2\left(1 + \frac{\alpha_{\rm MRT}^2 \eta \big| \hat{\bh}_k^H \bZ \hat{\bh}_k \big|^2}{ \alpha_{\rm MRT}^2 \eta \sum_{\substack{k'\in \cK \\ k'\neq k}} ||\hat{\bh}_k^H \bZ \hat{\bh}_{k'}||^2 + \frac{ \alpha_{\rm MRT}^2 \eta}{(1+\tau)(1+\mu_k)} \sum_{i=1}^{K}||\bZ \hat{\bh}_i||^2  +N_0} \right)}.
\end{align}

\end{figure*}
Along the same lines, we will consider the case with mutual coupling first and then extend to the case without mutual coupling, as the latter case is a special case of the former.

\subsubsection{With mutual coupling}
For notational convenience, denote $\breve{\bh}_k = \bLambda^{\frac{1}{2}} \bQ^T \hat{\bh}_k$.  We first introduce the following lemma.

\begin{lemma}\label{IP_MC_hk_hi_square_lemma}
The expectation for the same inner product of two same columns in $\breve{\bH}$ can be expressed as
  \begin{align}
    \expect{||\breve{\bh}_k||^2} = \expect{\breve{\bh}_k^H \breve{\bh}_k} = \trace{\bLambda^2}\frac{\mu_k(1+\tau)+\tau}{(1+\mu_k)(1+\tau)},
  \end{align}
and the expectation of the norm square of the inner product of any two different columns in $\hat{\bH}$ is
  \begin{align}\label{rate_MC_imCSI}
    &\expect{|\breve{\bh}_k^H \breve{\bh}_i|^2}\notag \\ = 
   & \begin{cases}
     \frac{1}{(1+\mu_k)^2}\left\{\trace{\bLambda}^2 \mu_k^2 + \left[2\trace{\bLambda}^2\mu_k + 2\mu_k \trace{\bLambda^2}\right]\frac{\tau}{1+\tau} \right. \\
       \vspace{3ex}
   \left. + \left[\trace{\bLambda}^2+\trace{\bLambda^2}\right](\frac{\tau}{1+\tau})^2\right\}, \qquad \qquad \qquad \,  k=i, \\ 
    \frac{1}{(1+\mu_k)(1+\mu_i)}\left[\mu_k \mu_i |\tilde{Q}(k,i)|^2+\trace{\bLambda^2}(\mu_i+\mu_k)\frac{\tau}{1+\tau} \right.\\
    \left. +\trace{\bLambda^2}(\frac{\tau}{1+\tau})^2\right], \qquad \qquad \qquad \qquad \qquad \: \: \:\: \: \: \, k \neq i.
    \end{cases}
  \end{align}
\end{lemma}
\begin{proof}
  See Appendix \ref{IP_MC_hk_hi_square_proof}.
\end{proof}

Utilizing \lemmref{IP_MC_hk_hi_square_lemma}, the ergodic SE of the $k$th user is shown in \eqref{IP_asy_SE_MC} at the top of next page,
\begin{figure*}
\begin{align}\label{IP_asy_SE_MC}
   R_{\rm MC}^{IP,k} \approx \tilde{R}_{\rm MC}^{IP,k}
  = \log_2\left( 1 + \frac{ \alpha_{\rm MRT}^2 \eta }{ (1+\mu_k)^2 }  \frac{\trace{\bLambda}^2 \left(\mu_k+\frac{\tau}{1+\tau}\right)^2 + \trace{\bLambda^2} \left(2\mu_k\frac{\tau}{1+\tau} +\left(\frac{\tau}{1+\tau}\right)^2 \right)}{ \tilde{f}_{MC} } \right).
\end{align}
\hrulefill
\end{figure*}
where $\tilde{f}_{MC}$ is given by
\begin{align}\label{tilde_f_MC}
 \tilde{f}_{MC} \triangleq & \frac{\alpha_{\rm MRT}^2  \eta}{1+\mu_{k}}\sum_{\substack{k'\in \cK \\ k'\neq k}}\left\{ \frac{\mu_k \mu_{k'}}{1+\mu_{k'}} |\tilde{Q}(k,k')|^2 \right. \notag \\
 &\left.+ \trace{\bLambda^2}\frac{\mu_{k'} +\mu_k}{1+\mu_{k'}}\frac{\tau}{1+\tau}+\frac{\trace{\bLambda^2}}{1+\mu_{k'}}(\frac{\tau}{1+\tau})^2\right\}\notag \\
 &+ \frac{\trace{\bLambda^2}\alpha_{\rm MRT}^2  \eta}{(1+\tau)(1+\mu_k)} \sum_{i=1}^{K}\frac{\mu_i(1+\tau)+\tau}{(1+\mu_i)(1+\tau)}+N_0.
\end{align}

\subsubsection{Without mutual coupling}
Substituting $\bZ = \bI$ and $\eta = 1$ into \eqref{IP_asy_SE_MC}, we have
\begin{align}\label{imCSI_rate_k_approx}
  &  R_{\rm nMC}^{IP,k}  \approx  \tilde{R}_{\rm nMC}^{IP,k} \notag \\
  & = \log_2\left(1 + \right. \notag \\
  &\left.\frac{ \frac{\alpha_{\rm MRT}^2  }{(1+\mu_k)^2}\left(N^2 (\mu_k + \frac{\tau}{1+\tau})^2 + N ( 2\mu_k + (\frac{\tau}{1+\tau})^2 ) \right)}{\tilde{f}} \right),
\end{align} 
where $\tilde{f}$ is 
\begin{align}\label{tilde_f}\
\tilde{f}  \triangleq &  \frac{\alpha_{\rm MRT}^2  \eta}{1+\mu_{k} }\sum_{\substack{k'\in \cK \\ k'\neq k}}\frac{1}{1+\mu_{k'}} \left\{ \mu_k \mu_{k'} \left|Q(k,i)\right|^2 \right.  \notag \\
& \left. +N(\mu_{k'} +\mu_k)\frac{\tau}{1+\tau}+N(\frac{\tau}{1+\tau})^2\right\} \notag \\
& + \frac{N\alpha_{\rm MRT}^2  \eta}{(1+\tau)(1+\mu_k)} \sum_{i=1}^{K}\frac{\mu_i(1+\tau)+\tau}{(1+\mu_i)(1+\tau)}  +N_0.
\end{align}

Under imperfect CSI, the system performance is further affected by channel estimation errors. To gain more insights, we will analyze two special cases of \eqref{IP_asy_SE_MC} and \eqref{imCSI_rate_k_approx} in the following sections.

\subsubsection{Achievable SE under Pure LoS and Rayleigh Fading Scenarios} Unlike the case with perfect CSI, the system SE under imperfect CSI is influenced not only by background noise and inter-user interference but also by channel estimation errors. The relative impact of these factors varies naturally with different Ricean factors. In a pure LoS scenario, channel estimation errors have no impact. However, as the Ricean factor decreases, the influence of channel estimation errors on system SE becomes increasingly significant. To better illustrate the impact of channel estimation errors on the system SE as the number of antennas $N$ increases, we will primarily focus on the Rayleigh fading scenario in the subsequent analysis, which represents an extreme case of Ricean propagation.

When $\mu_k = 0, \forall k\in\cK$, the asymptotic ergodic SEs of the $k$th user in \eqref{IP_asy_SE_MC} and \eqref{imCSI_rate_k_approx} reduce to \eqref{IP_Ray_rate_k_MC} below and \eqref{rate_k_Ray_nMC_imCSI} at the top of next page, respectively,
\begin{subequations}\label{IP_Ray_rate_k_MC}
\begin{align}
  &\tilde{R}_{\rm MC,Ray}^{IP,k} \notag \\
  &= \log_2\left( 1+\frac{\alpha^2_{\rm MRT} (\frac{\trace{\bLambda}^2}{\trace{\bLambda^2} } +1)}{\alpha^2_{\rm MRT}(K-1) + \frac{\alpha^2_{\rm MRT} K }{\tau} +\frac{N_0}{\trace{\bLambda^2} \eta} (1+\frac{1}{\tau})^2 }\right) \\
  &\overset{(a)}{\approx} \log_2\left( 1+\frac{\alpha^2_{\rm MRT} (N +1)}{\alpha^2_{\rm MRT}(K-1) + \frac{\alpha^2_{\rm MRT} K }{\tau} +\frac{N_0}{\trace{\bLambda^2} \eta} (1+\frac{1}{\tau})^2 }\right)\!,
\end{align}
\end{subequations}
where $(a)$ comes from the approximation of $\frac{\trace{\bLambda}^2}{\trace{\bLambda^2}} \approx N$, as mentioned in \eqref{ratio_lambda}.
\begin{figure*}[t]
\begin{align}\label{rate_k_Ray_nMC_imCSI}
 \tilde{R}_{\rm nMC,Ray}^{IP,k} =  \log_2 \left( 1 +\frac{\alpha^2_{\rm MRT} (N+1)}{\alpha^2_{\rm MRT}(K-1) + \frac{\alpha^2_{\rm MRT} K}{\tau} + \frac{N_0}{N}(1+\frac{1}{\tau})^2}\right).
\end{align}
\hrulefill
\end{figure*}

\begin{remark}
With high transmit power, the background noise $N_0$ can be ignored. Although $\tilde{R}_{\rm MC,Ray}^{IP,k}$ will be slightly lower than $\tilde{R}_{\rm nMC,Ray}^{IP,k}$,  they will show minimal differences. It is because the ratio $\trace{\bLambda}^2/\trace{\bLambda^2}$ is always smaller than $N$, but as shown in \figref{fig_ratio}, the difference is negligible.
In other words, the SE of a continuous array system will serve as the upper bound of the system SE. 
On the contrary, if the background noise $N_0$ cannot be ignored and the radiation efficiency $\eta$ is acceptable, which is possible when the number of antenna elements is moderate, the SE considering mutual coupling will potentially outperform the SE without considering mutual coupling. This is because $\trace{\bLambda^2} \gg N$ as well as $\trace{\bLambda^2}\eta \geq N$, and thus, the existence of mutual coupling will mitigate the effect of $\tau$ on the term dependent on $N_0$ more prominently than $N$, leading to a better system SE. 
However, when $N$ continuously increases, $\eta$ will become too small to reduce the system' SE.
In this case, the SE without mutual coupling will become better than the SE with mutual coupling.  Note that in the extreme case of $N \rightarrow \infty$, $\tilde{R}_{\rm MC,Ray}^{IP,k}$ will converge to $\tilde{R}_{\rm nMC,Ray}^{IP,k}$, which is consistent with \remref{remark_los}. 
\end{remark}

\section{Numerical Results}\label{num_res}

In this section, we provide simulation results to verify the obtained theoretical analysis.
In our simulations, we set the frequency of carrier wavelength as 28 GH. Note that we have not imposed any specific assumptions on the signal frequency, indicating that the frequency parameter does not affect our analytical conclusions. The physical size of the antenna array at the BS is $15\lambda \times 15 \lambda$.
There are $K = 10$ single-antenna users distributed randomly and uniformly within a circular ring centred at the BS, with a radius ranging from 5 to 10 m.  Note that the distance exceeds the uniform power distance, which is $\sqrt{\frac{\Gamma^{2/3}}{1-\Gamma^{2/3}}}\frac{D}{2} \approx 0.5$ m with $\Gamma = 90\%$ and $D = 15 \sqrt{2} \lambda$ \cite{9617121}. Thus, we can accurately consider uniform large-scale path loss, as mentioned in \subsecref{channel model}, i.e., $\forall n_y,n_z,$ $\beta_k^{(n_y,n_z)}  = \beta_k$ and $\tilde{\beta}_{k}^{(n_y,n_z)} = \tilde{\beta}_k$.
The large-scale path loss model is given as follows \cite{10579914}:
\begin{subequations}
\begin{align}
  \beta_k \text{[dB]} = 61.4 + 20\log_{10}(r_k),\\
  \tilde{\beta}_k \text{[dB]} = 72 + 29.2\log_{10}(r_k).
\end{align}
\end{subequations}
For convenience, we assume that each user has the same Ricean factor. If not explicitly stated, the parameters will be set as follows: The Ricean factor is $\mu_k = 10$ dB, $\forall k \in \cK$; the length of pilots is $ p = 10$ ($p \geq K$) for the imperfect CSI case; the variance of noise $N_0 = -94$ dBm; the loss factor is set to be $\gamma = 0.1$. The area of the transmitting array is assumed to be fixed, which means that decreasing the antenna spacing will allow more antennas to be deployed in this array. For all numerical results in this section, we will denote Monte-Carlo results in \eqref{expectations_SE} and asymptotic results in \eqref{rate_k_MRT_Ricean}, \eqref{IP_asy_SE_MC} as ``Sim'' and  ``Asy'', respectively. The primary comparison scheme for all simulations is the case without mutual coupling, while in each subsection, we will introduce the corresponding benchmarks to demonstrate the advantages of near-field propagation and the MRT scheme.

\subsection{Impact of Antenna Spacing}
\begin{figure}[htbp]
\captionsetup{font={small}}
    \centering
    \includegraphics[width = 0.42\textwidth]{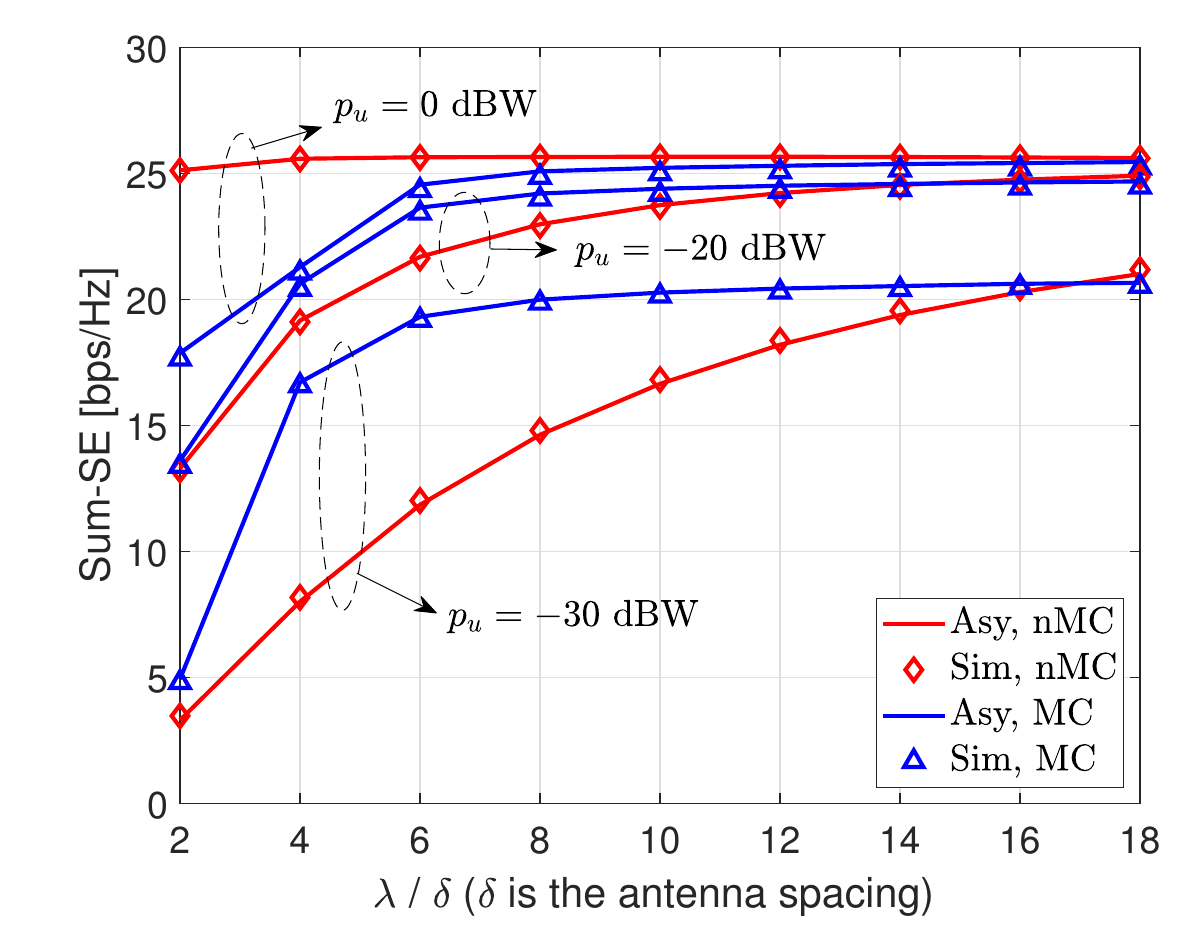}
    \caption{Sum-SE versus different antenna spacing.}
    \label{fig_impact_dense_array}
\end{figure}
In this subsection, we investigate the impact of antenna spacing on the achieved SE under perfect CSI.
Figure \ref{fig_impact_dense_array} illustrates the sum-SE of the considered multi-user HMIMO system for the cases of $p_u = 0$ dBW, $p_u = -20$ dBW and $p_u = -30$ dBW. The solid lines represent the case of asymptotic results, while the dotted lines represent the simulated results.
First, it can be observed that the analytical results match well with the simulation results, which confirm the correctness of our theoretical analysis. 
In the high transmit power regime ($p_u = 0$ dBW), the system SE with mutual coupling is always lower than that without mutual coupling; however, the gap between them gradually decreases as the number of antenna array elements increases.
In the low transmit power regime ($p_u = -20,-30$ dBW), the system SE with mutual coupling outperforms that without mutual coupling; by increasing $N$, the gap between them also gradually decreases.  In other words, the system SE without mutual coupling represents the upper bound of the SE when $N$ tends to infinity. 
The analysis above is consistent with \remref{remark_ray} and \remref{remark_los}.

\subsection{Impact of Near-field Channel Model}
\begin{figure}[htbp]
\captionsetup{font={small}}
    \centering
    \includegraphics[width = 0.42\textwidth]{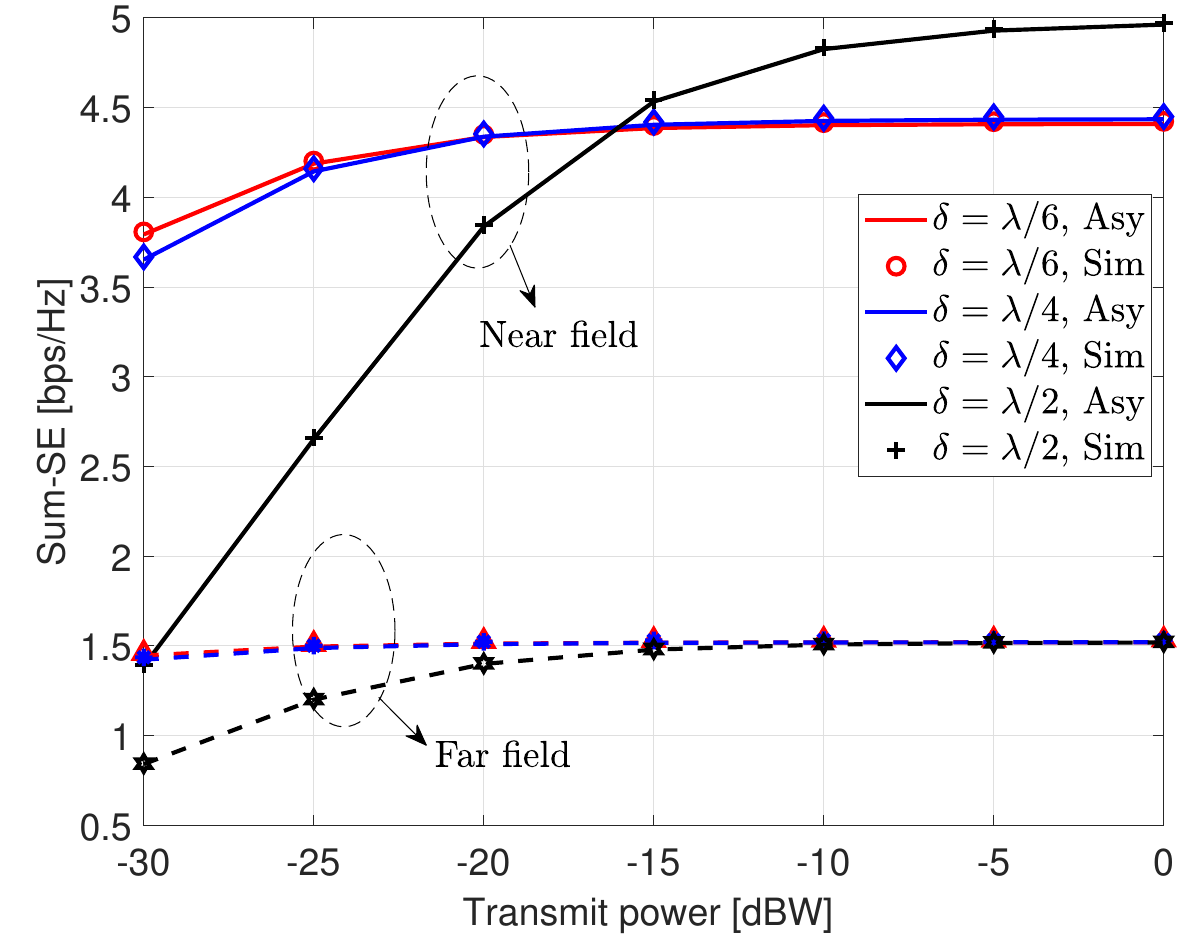}
    \caption{Sum-SE with different channel models.}
    \label{near_far}
\end{figure}
\begin{figure}[htbp]
    \captionsetup{font={small}}
    \centering
    \includegraphics[width = 0.42\textwidth]{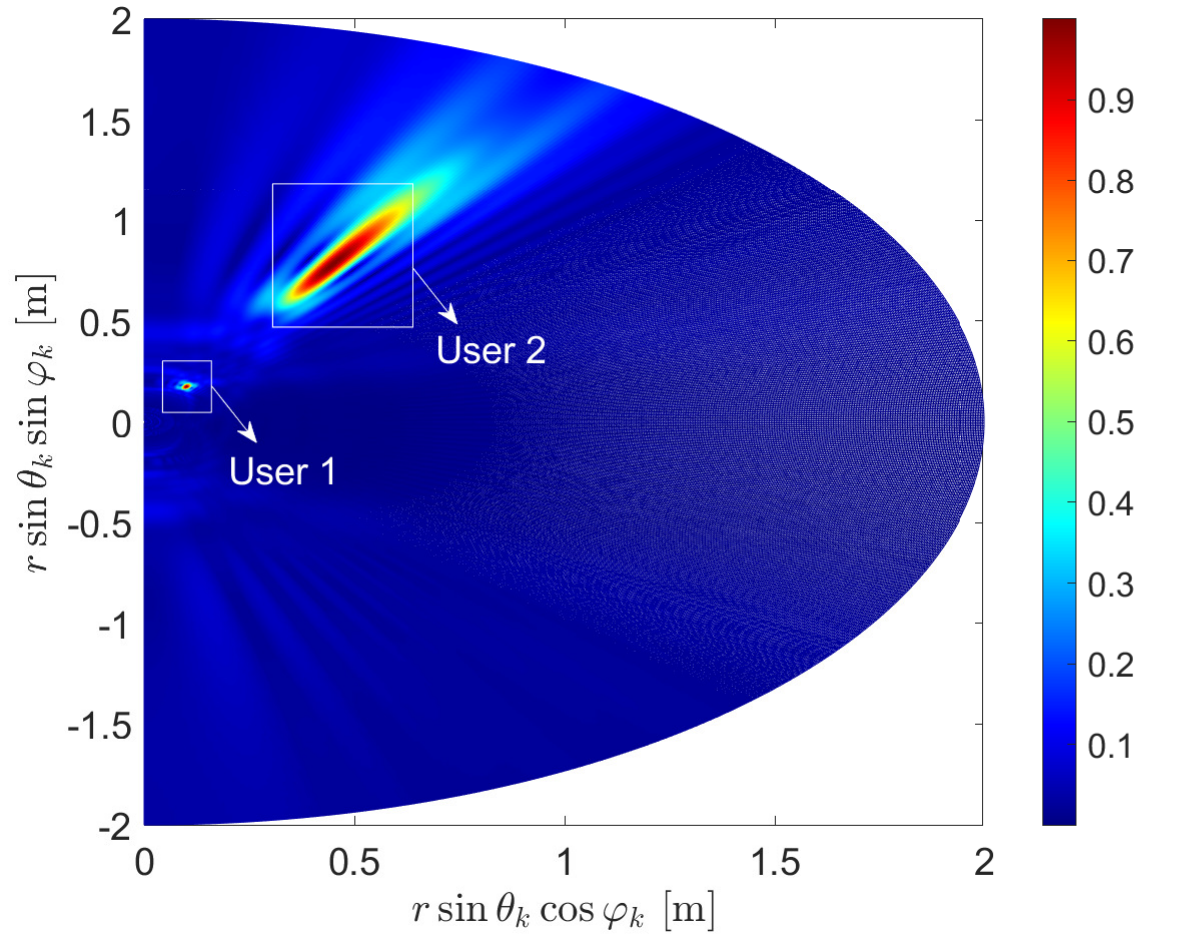}
    \caption{The beam pattern of two users in near field.}
    \label{beampattern}
\end{figure}
In this subsection, we compare the  sum-SE achieved for the near-field and far-field channel models in a pure LoS scenario, where the inter-user interference is mostly severe. According to the planar-wave based far-field model, the prorogation distance between  the $(n_y,n_z)$th antenna and user $k$ is
\begin{align}
  a^{\left(n_y, n_z\right)}\left(\theta_{k}, \varphi_{k}, r_{k}\right) = e^{-\jmath\frac{2\pi}{\lambda}r_k}\exp \!\left\{\!-\jmath \frac{2 \pi}{\lambda}\left(r_{k}^{\left(n_y, n_z\right)}-r_{k}\right)\!\right\},
\end{align}
where the additional distance $\tilde{\Delta} r_{k}^{\left(n_y, n_z\right)} \triangleq r_{k}^{\left(n_y, n_z\right)}-r_{k}$ is
\begin{align}
 \tilde{\Delta} r_{k}^{\left(n_y, n_z\right)} =  \delta \left(n_y \sin \theta_{k} \sin \varphi_{k} + n_z \cos \theta_{k}\right).
\end{align}
It can be observed that the far-field channel model only considers the angles' terms, while the near-field channel model in \eqref{near-field array response vector} considers the quadratic distance term $\delta^2 \left(n_y^2+n_z^2\right)/(2 r_{k})$ additionally.
The users' angles are set to be the same, which are $(\theta_k,\varphi_k) = (\pi/2, 2\pi/3), \forall k$, while the distance of each user is set randomly and within the ranges of $(0,1]$ m.
``Far-field'' in \figref{near_far} employs the conventional far-field channel model, which only considers differences in users' angular positions, and therefore leads to the most severe inter-user interference, i.e., the worst case. It can be observed that densifying the antenna array will enhance the system SE slightly in the far-field case. Compared with the ``Far-field'' case, the system SE experiences a significant improvement under the near-field channel model. This improvement is attributed to the inclusion of the distance term in the near-field channel, which enhances the DoF for distinguishing the phases of different users. The above phenomenon is consistent with \remref{near_far_remark}. 

To further reveal the performance gains brought by the near-field propagation, we provide the beam pattern in \figref{beampattern}, which helps us to understand why the sum-SE in the worst case is always larger with the near-field channel model than that with the far-field channel model.
Figure \ref{beampattern} shows a two users' beam pattern in the near field, with the position of user 1 being $(r,\theta,\varphi) = (0.2,\pi/2,\pi/3)$ and the position of user 2 being $(r,\theta,\varphi) = (0.9,\pi/2,\pi/3)$. It can be observed that the users' beam patterns become maximal at the same angle but at different distances,
indicating that a precise beamfocusing can be realized with the near-field channel, since it can mitigate the inter-user interference in the worst case even with MRT precoding.

\subsection{Impact of Ricean Factor}
\begin{figure}[htbp]
\captionsetup{font={small}}
    \centering
    \includegraphics[width = 0.42\textwidth]{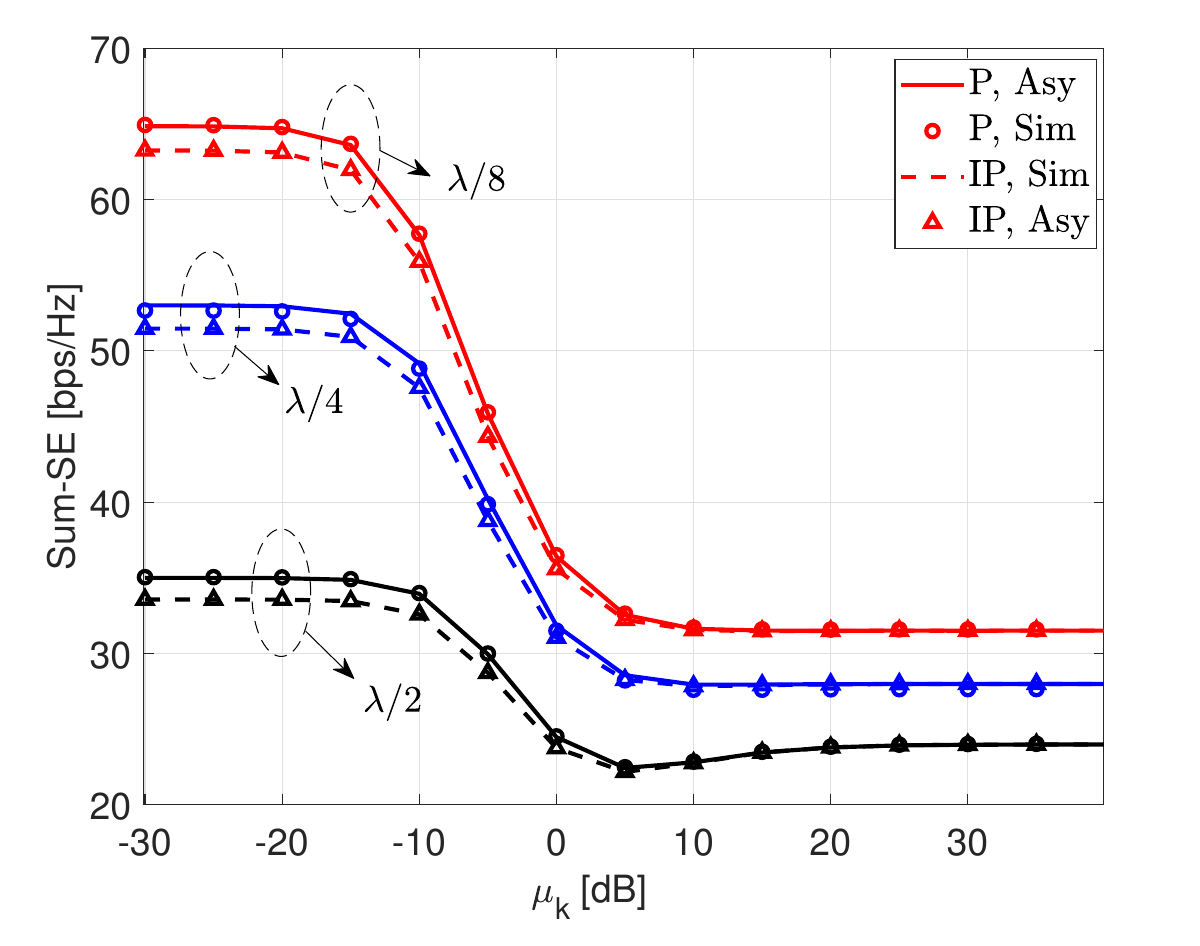}
    \caption{Sum-SE versus different Ricean factor.}
    \label{K_factor_SE}
\end{figure}
In this subsection, we illustrate the impact of Ricean factor on the system SE. Note that we randomly generate 100 times different users' distributions, and the system SE shown in \figref{K_factor_SE} represents the average SE.
In \figref{K_factor_SE}, we can first observe that when the system transits from a NLoS-dominated propagation scenario to a LoS-dominated scenario,  MRT leads to severe inter-user interference, resulting in the degraded sum-SE. Second, although reducing the antenna spacing leads to an increase in the system SE, the rate of such increase diminishes as the antenna spacing decreases from $\lambda/2$ to $\lambda/6$.
Besides, when the Ricean factor increases, the sum-SE under imperfect CSI will have a slight increase at $\lambda/2$ antenna spacing.
This is because the impact of inter-user interference is less significant than the impact of channel estimation errors at $\lambda/2$ antenna spacing.  Increasing the Ricean factor reduces the impact of channel estimation error on the system SE, thereby leading to a slight improvement in it.
However, as the antenna spacing decreases from $\lambda/2$ to $\lambda/6$ or even smaller, the inter-user interference in the system becomes increasingly severe due to the impact of mutual coupling on the transmitted signals of each user. In this case, although increasing the Ricean factor can reduce the impact of channel estimation error on the system SE, inter-user interference becomes the dominant source of degradation. This interference limits the growth of the system SE.
In fact, due to the limitations of MRT, the sum-SE in pure LoS scenarios is heavily influenced by the inter-user interference. As a result, the relative performance between the NLoS sum-SE and the LoS sum-SE can vary depending on the users' positions, and that is why we choose to average the system' SE over the users' distributions.

\subsection{Impact of Loss Factor}
\begin{figure}[htbp]
    \captionsetup{font={small}}
    \centering
    \includegraphics[width = 0.42\textwidth]{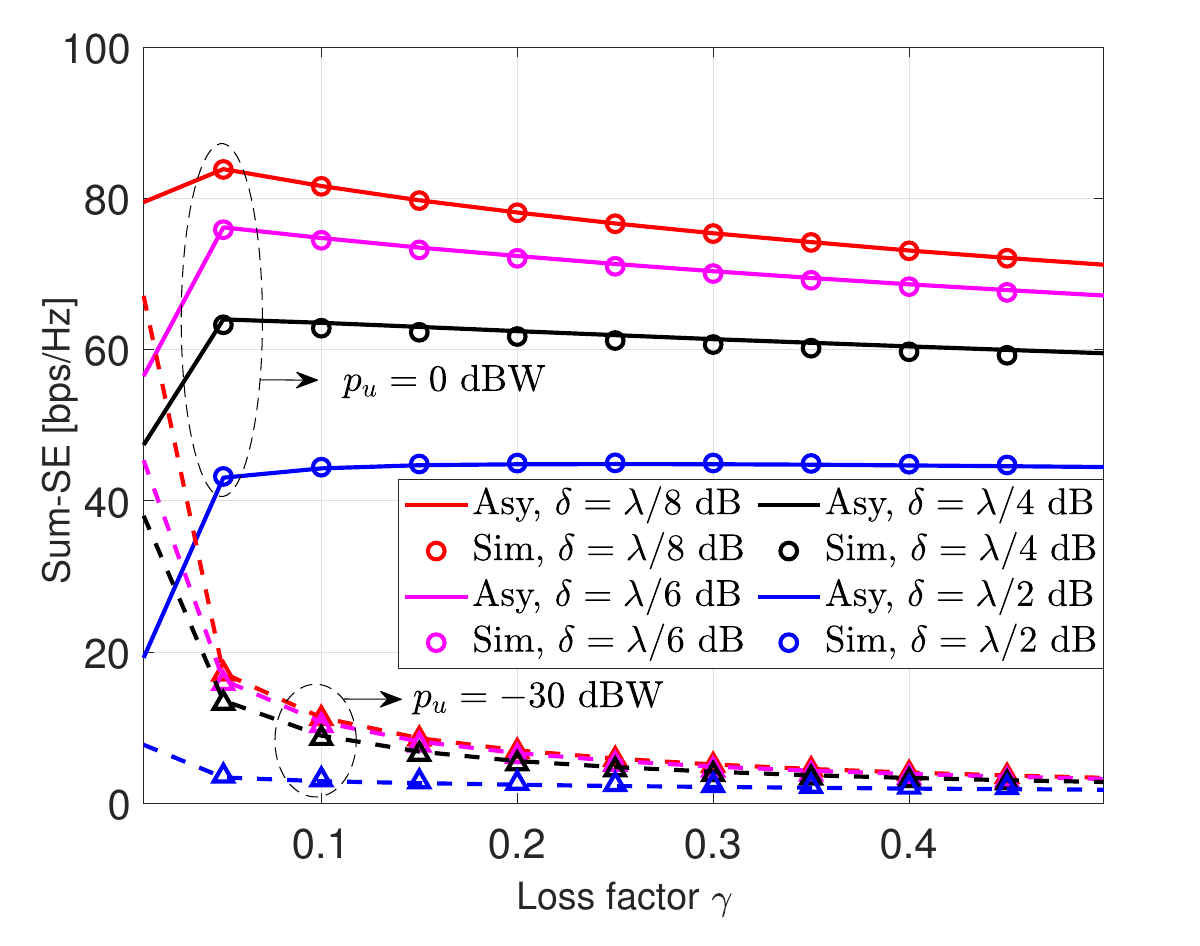}
    \caption{Sum-SE versus different loss factor.}
    \label{loss_factor}
\end{figure}
In this subsection, we compare the impact of the loss factor on the system SE. 
Figure \ref{loss_factor} illustrates the variation in the system SE by increasing loss factor under two scenarios: $p_u =0 $ dBW and $p_u = -30$ dBW.  The effect of loss factor on the system SE differs under high and low transmit power constraints.
Under high transmit power conditions ($p_u = 0$ dBW), the primary noise source is the inter-user interference. In this case, as the loss factor increases, the effective signal will increase slightly at first, leading to a temporary increase in the system SE. However, as the loss factor continues to increase, the system SE begins to reduce. This is due to the rapid increase in the inter-user interference within the system, which hinders the improvement on the system SE. This reduction becomes more pronounced as the antenna spacing decreases.
Under low transmit power condition ($p_u = -30$ dBW), the primary noise source becomes the background noise $N_0$. In this case, the useful signal can be approximately considered invariant, while an increase in the loss factor amplifies the impact of background noise on the system SE. Consequently, increasing the loss factor indirectly decreases the system SE.

\subsection{Impact of Imperfect CSI}
\begin{figure}[htbp]
    \captionsetup{font={small}}
    \centering 
    \subfigure[$p_u = 0$ dBW (relatively large transmit power)]{
    \includegraphics[width = 0.42\textwidth]{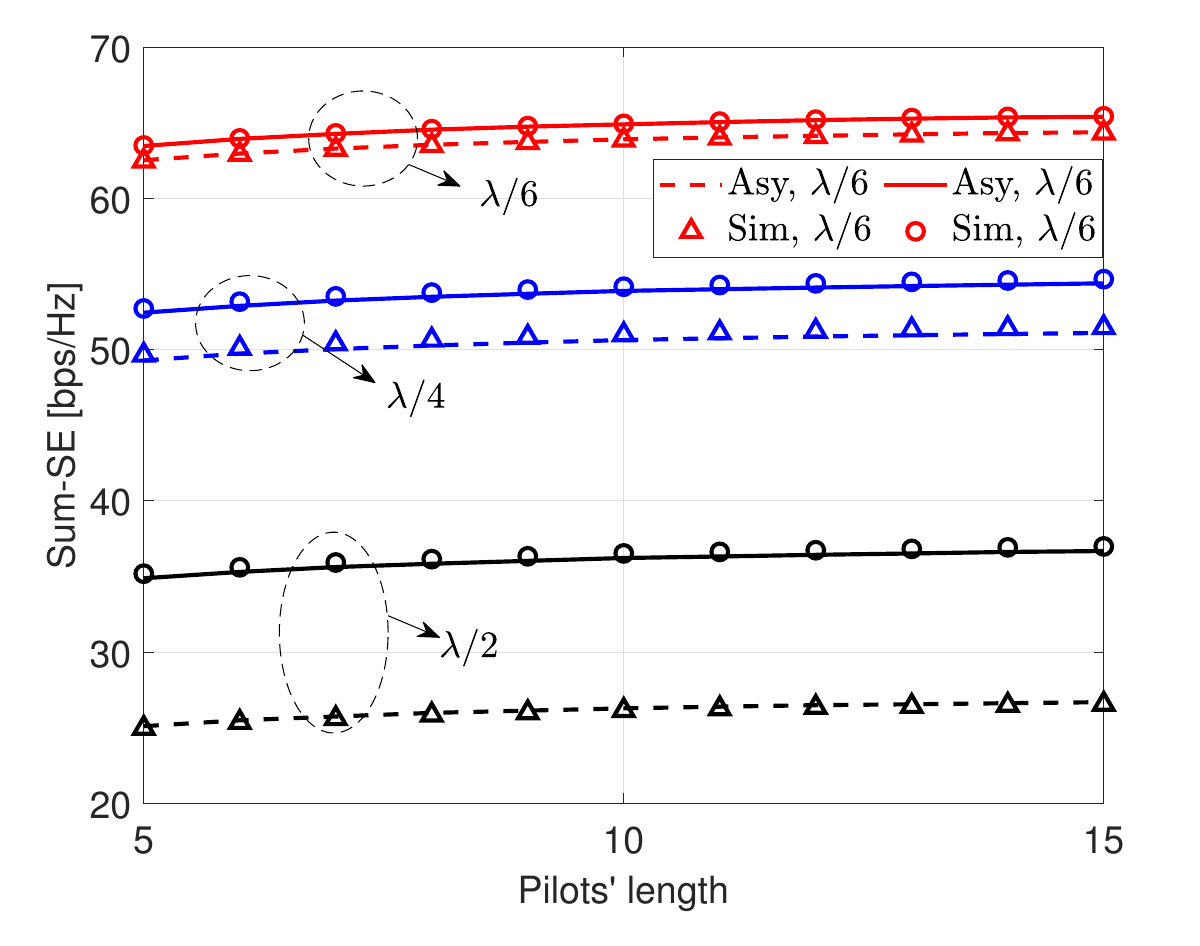}\label{large_pu_imCSI}
     }
     \subfigure[$p_u = -20$ dBW (relatively small transmit power)]{
    \includegraphics[width = 0.42\textwidth]{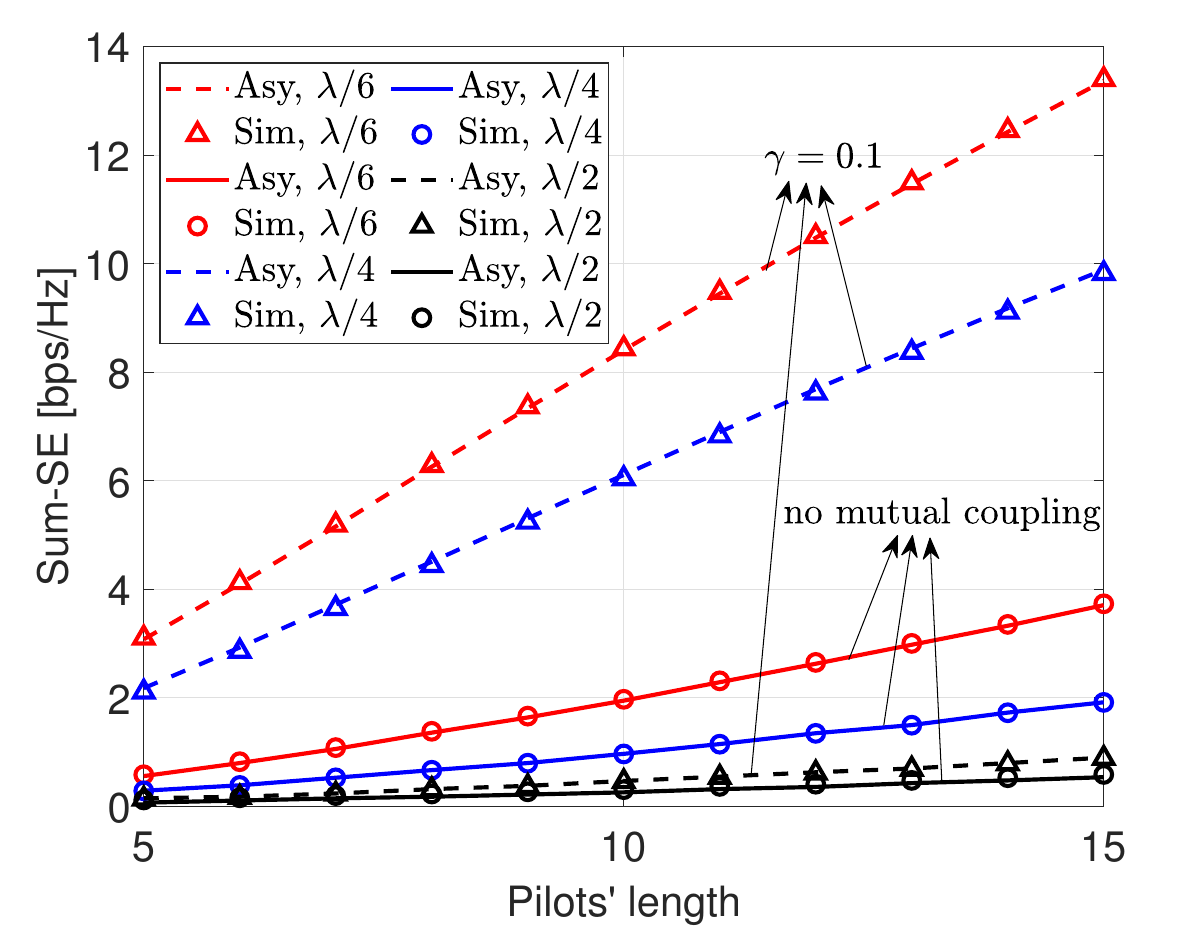}\label{small_pu_imCSI}
     }
    \caption{Sum-SE with imperfect CSI.}
    \label{IP_SE}
\end{figure}
In this subsection, we investigate the impact of channel estimation errors on the system sum-SE.
To better illustrate the difference in sum-SE between systems with and without mutual coupling under imperfect CSI, we set $\mu_k = 0$ here, fully considering the Rayleigh fading scenario.
We compare the system SE under imperfect CSI with different antenna spacings (i.e., different mutual coupling levels) as a function of the pilot length, which also corresponds to pilot energy in this context. From \figref{IP_SE}, we can see that there are two ways to reduce the impact of channel estimation errors on the system SE by increasing the pilot energy or increasing the number of array antenna elements. 
As shown in the previous analysis and \figref{large_pu_imCSI}, when $N$ increases to a certain level and the transmit power is relatively large compared with the background noise $N_0$, the SE with mutual coupling becomes almost identical to that without considering mutual coupling, with the former being slightly lower than the latter in practice.  It is also clear that increasing the pilot energy brings almost no benefit to the growth of the system SE when the transmit power is relatively high. However, increasing the level of array densification can still bring gains to the system SE under such circumstances.
On the contrary, if the system's transmit power is relatively low compared to the background noise $N_0$, as shown in \figref{small_pu_imCSI}, the SE with mutual coupling will be larger than that without mutual coupling when $N$ is not large enough. This is because the presence of mutual coupling mitigates the impact of $N_0$, specifically reflected in the term $N_0/\trace{\bLambda^2}$ in the denominator of the SE of user $k$.

\subsection{Impact of Precoding Methods}
\begin{figure}[htbp]
    \captionsetup{font={small}}
    \centering
    \includegraphics[width = 0.42\textwidth]{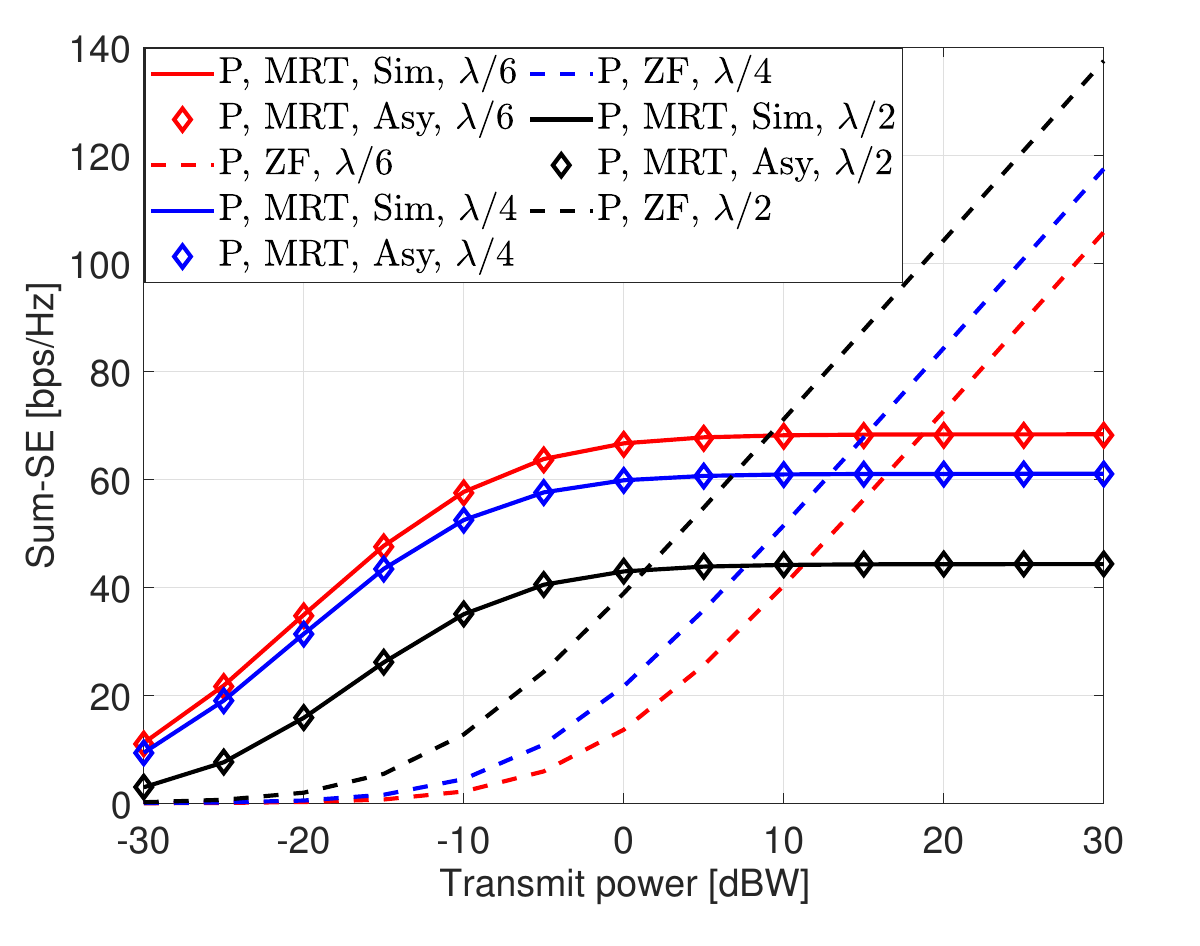}
    \caption{Sum-SE with different precoding methods.}
    \label{ZF_MRT}
\end{figure}
In the previous analysis and simulations, it has been demonstrated that inter-user interference significantly impacts the system SE.
A relatively simple yet classic approach is to utilize the zero-forcing (ZF) precoding method to mitigate the inter-user interference, especially in the high transmit power regime. This subsection compares the impact of ZF precoding and MRT on the system SE with perfect CSI. Here, the mutual coupling matrix is incorporated into the ZF precoding design, and the resulting precoding matrix is given as: 
\begin{align}
  \bW_d = \alpha_{ZF}\bZ^{-1}\left(\bH \bH^H\right)^{-1}\bH.
\end{align}
Figure \ref{ZF_MRT} illustrates the impact of these two precoding schemes on the system SE. It can be observed that 
in the low transmit power regime, MRT and smaller antenna spacing (which leads to severe mutual coupling) can be selected to enhance the system SE. In contrast, when the transmit power is relatively high, the ZF precoding should be adopted. 
However, it is shown that reducing the antenna element spacing will decrease the effective radiation efficiency $\eta$ in this scenario, thereby reducing the strength of the useful signal and thus, leading to the reduction of the system SE. Note that the ZF performance presented here represents only an ideal case, where the inter-user interference is totally eliminated by incorporating the mutual coupling matrix into ZF precoding. It may not be entirely practical, as the mutual coupling matrix might be more complex in practice. This highlights that interference cancellation among users in holographic communication is a critical issue, which will motivate our future research.

\section{Conclusion}\label{conclusion}
This paper investigated the SE of a near-field multi-user downlink HMIMO system over Ricean fading channels
with both perfect and imperfect CSI.
For perfect CSI,  we derived the system SE for the cases with and without mutual coupling, respectively. It was shown that the SE with a continuous array was the upper bound of the system. However, for a moderate number of antenna elements, it was possible that the SE with mutual coupling might outperform that without mutual coupling, especially in the low transmit power regime. Meanwhile,
as the Ricean factor increased, inter-user interference became the primary source of the system's interference. The near-field channel provided distance-domain DoF, which helped to distinguish users and thereby reduced the occurrence of worst-case scenarios.
For imperfect CSI, we considered the channel estimation errors and also derived the system' SE under MRT for the cases with and without mutual coupling.
It was unveiled that the SE with a continuous array was still the upper bound of the system.  In the low transmit power regime, the system SE could be improved by increasing the pilot power and the antenna element density. In the high transmit power regime, increasing the pilot power had a limited effect on enhancing the system SE. However, increasing the antenna element density remained highly beneficial for improving the system SE. 
Last but not least, we compared the impact of ZF precoding and MRT on the system SE through simulations, which confirmed the necessity of investigating the inter-user interference in future work.

\appendices

\section{Proof of Lemma 1}\label{hk_hi_norm_square_lemma}
In the following proof, we will ignore the large-scale path-loss coefficients $\beta_k$ and $\tilde{\beta}_k$, as they are deterministic.

  When $k = i$, 
\begin{align}
 \expect{\tilde{\bh}_k^H \tilde{\bh}_k} = \expect{\trace{\bh_k^H \bLambda \bh_k}} = \trace{\bLambda},
\end{align}
while $||\tilde{\bh}_k||^4$ is given in \eqref{tilde_h_4} at the top of next page,
\begin{figure*}[!t]
\begin{align}\label{tilde_h_4}
 &||\tilde{\bh}_k||^4  = \left(\frac{\mu_k}{1+\mu_k}\right)^2 \trace{\bLambda}^2+ \frac{1}{(1+\mu_k)^2} \!\left[\sum_{n=1}^{N}\lambda_n(s_{nk}^2+t_{nk}^2)\right]^2 + \frac{4 \mu_k \sqrt{\mu_k} }{(1+\mu_k)^2} \trace{\bLambda} \sum_{n=1}^{N}\lambda_n (\rho_{n k}^{c}s_{nk} -\rho_{nk}^{s}t_{nk} ) \notag \\
 &+ \frac{2  \mu_k }{(1+\mu_k)^2} \trace{\bLambda}\sum_{n=1}^{N}  \lambda_{n}(s_{n k}^2 + t_{nk}^2) + \frac{4 \mu_k }{(1+\mu_k)^2} \left[ \sum_{n=1}^{N}\lambda_n( \rho_{nk}^{c}s_{nk} -\rho_{nk}^{s}t_{nk}  ) \right]^2 \notag \\
  &+\frac{4  \sqrt{\mu_k} }{(1+\mu_k)^2} \sum_{n_1,n_2}^{N}  \lambda_{n_1} \lambda_{n_2} (\rho_{n_1 k}^{c}s_{n_1 k} -\rho_{n_1 k}^{s}t_{n_1 k} )(s_{n_2k}^2 + t_{n_2k}^2).
\end{align}
\hrulefill
\end{figure*}
where $s_{nk}$ and $t_{nk}$ represent the independent real and imaginary parts of $[\bh_{k, NLoS}]_{n}$, respectively, while $\rho_{nk}^{c}$ and $\rho_{nk}^{s}$ represent the independent real and imaginary parts of $[\bh_{k, LoS}]_{n}$, respectively.

After removing the terms with zero expectation, $\expect{||\tilde{\bh}_k||^4}$ is given by
\begin{subequations}\label{h_k_4_MC}
  \begin{align}
  \expect{||\tilde{\bh}_k||^4} & =  (\frac{\mu_k}{1+\mu_k})^2 \trace{\bLambda}^2 + \frac{2 \mu_k}{(1+\mu_k)^2} \trace{\bLambda}^2 \notag \\
  & + \frac{2\mu_k}{(1+\mu_k)^2}\trace{\bLambda^2} + \frac{\trace{\bLambda}^2+\trace{\bLambda^2}}{(1+\mu_k)^2} \\
  & =  \trace{\bLambda}^2+ \frac{(2\mu_k+1)\trace{\bLambda^2}}{(1+\mu_k)^2}.
   \end{align}
\end{subequations}

When $k\neq i$, we obtain the real and imaginary parts of $\bh_k^H \bh_i$ first, which are
\begin{subequations}\label{real_imag_hk_hi}
  \begin{align}
&\real{\bh_k^H \bh_i} \notag \\
 & =  \frac{1}{ \sqrt{\left(\mu_k+1\right)\left(\mu_i+1\right)} } \left\{\sum_{n=1}^N \lambda_n \left[ \sqrt{\mu_k} \left(\rho_{nk}^c s_{ni} - \rho_{nk}^{s} t_{ni} \right)  \right. \right. \notag \\
&\left. \left. +\sqrt{\mu_i} \left(\rho_{ni}^c s_{nk} - \rho_{ni}^{s} t_{nk} \right) + \left(s_{nk}s_{ni}+t_{nk}t_{ni} \right)  \right]   \right.\notag \\
&\left. +\sqrt{\mu_k \mu_i} \real{\tilde{Q}(k,i)}    \right\},\\
&\imag{\bh_k^H \bh_i} \notag \\
 & =  \frac{1}{ \sqrt{\left(\mu_k+1\right)\left(\mu_i+1\right)}} \left\{\sum_{n=1}^N \lambda_n \left[ \sqrt{\mu_k} \left(\rho_{nk}^c t_{ni} + \rho_{nk}^{s} s_{ni} \right) \right. \right. \notag \\
 &\left. \left. - \sqrt{\mu_i} \left(\rho_{ni}^c t_{nk} + \rho_{ni}^{s} s_{nk} \right) + \left(s_{nk}t_{ni}-t_{nk}s_{ni} \right)  \right] \right.\notag \\
 &\left. +  \sqrt{\mu_k \mu_i} \imag{\tilde{Q}(k,i)} \right\}.
  \end{align}
\end{subequations}
Then, $\expect{|\bh_k^H \bh_i|^2}$ is 
\begin{subequations}
\begin{align}\label{hk_hi_MC}
  \expect{|\bh_k^H \bh_i|^2} &=  \expect{\real{\bh_k^H \bh_i}^2 + \imag{\bh_k^H \bh_i}^2 } \\
  &=  \frac{\mu_k\mu_i  \big|\tilde{Q}(k,i)\big|^2  + \trace{\bLambda^2}(\mu_k+\mu_i + 1) }{(\mu_k+1)(\mu_i+1)}.
\end{align}
\end{subequations}

\section{Proof of Lemma 2}\label{IP_MC_hk_hi_square_proof}
Similarly, we will also ignore the large-scale path-loss coefficients $\beta_k$ and $\tilde{\beta}_k$ in the following proof.

 When $k = i $, 
\begin{align}
  \expect{\breve{\bh}_k^H \breve{\bh}_k} = \trace{\bLambda^2}\frac{\mu_k(1+\tau)+\tau}{(1+\mu_k)(1+\tau)}.
\end{align}

When $k \neq i$, the proof is similar to that in Appendix \ref{hk_hi_norm_square_lemma}, as we still need to denote the real and imaginary parts of $\hat{\bh}_k^H\hat{\bh}_i$, and then square them to obtain the final results. The process is omitted for the sake of brevity.

\bibliographystyle{IEEEtran.bst}
\bibliography{Refabrv_20180802,ref_SE.bib}
	
\end{document}